\documentclass{article}

\usepackage{arxiv}

\usepackage[utf8]{inputenc} 
\usepackage[T1]{fontenc}    
\usepackage[hidelinks]{hyperref}       
\usepackage{url}            
\usepackage{booktabs}       
\usepackage{amsfonts}       
\usepackage{nicefrac}       
\usepackage{microtype}      
\usepackage{cite}
\usepackage{graphicx}
\usepackage{subfig}
\usepackage{array}

\newcolumntype{L}[1]{>{\raggedright\let\newline\\\arraybackslash\hspace{0pt}}m{#1}}
\newcolumntype{C}[1]{>{\centering\let\newline\\\arraybackslash\hspace{0pt}}m{#1}}
\newcolumntype{R}[1]{>{\raggedleft\let\newline\\\arraybackslash\hspace{0pt}}m{#1}}

\newcommand{\field}[1]{{\bf {#1}}}

\title{Estimation of COVID-19 under-reporting in \\Brazilian States through SARI} 

\author{
	Balthazar~Paixão\\
	CEFET/RJ\\
	\scriptsize{\texttt{balthazar.paixao@eic.cefet-rj.br}} \\
	\And
	Lais~Baroni\\
	CEFET/RJ\\
	\scriptsize{\texttt{lais.baroni@eic.cefet-rj.br}} \\
	\And
	Rebecca~Salles\\
	CEFET/RJ\\
	\scriptsize{\texttt{rebeccapsalles@acm.org}} \\
	\And
	Luciana~Escobar\\
	CEFET/RJ\\
	\scriptsize{\texttt{luciana.vignoli@eic.cefet-rj.br}} \\
	\AND
	Carlos~de~Sousa\\
	FIOCRUZ\\
	\scriptsize{\texttt{carlos.sousa@bio.fiocruz.br}} \\
	\And
	Marcel~Pedroso\\
	FIOCRUZ\\
	\scriptsize{\texttt{marcel.pedroso@icict.fiocruz.br}} \\
	\And
	Raphael~Saldanha\\
	FIOCRUZ\\
	\scriptsize{\texttt{raphael.saldanha@icict.fiocruz.br}} \\
	\AND
	Rafaelli~Coutinho\\
	CEFET/RJ\\
	\scriptsize{\texttt{rafaelli.coutinho@cefet-rj.br}} \\
	\And
	Fabio~Porto\\
	LNCC\\
	\scriptsize{\texttt{fporto@lncc.br}} \\
	\And
	Eduardo~Ogasawara\\
	CEFET/RJ\\
	\scriptsize{\texttt{eogasawara@ieee.org}} \\
}

\begin{document}
\maketitle

\begin{abstract}
Due to its impact, COVID-19 has been stressing the academy to search for curing, mitigating, or controlling it. However, when it comes to controlling, there are still few studies focused on under-reporting estimates. It is believed that under-reporting is a relevant factor in determining the actual mortality rate and, if not considered, can cause significant misinformation. Therefore, the objective of this work is to estimate the under-reporting of cases and deaths of COVID-19 in Brazilian states using data from the Infogripe on notification of Severe Acute Respiratory Infection (SARI). The methodology is based on the concepts of inertia and the use of event detection techniques to study the time series of hospitalized SARI cases. The estimate of real cases of the disease, called novelty, is calculated by comparing the difference in SARI cases in 2020 (after COVID-19) with the total expected cases in recent years (2016 to 2019) derived from a seasonal exponential moving average. The results show that under-reporting rates vary significantly between states and that there are no general patterns for states in the same region in Brazil.

The published version of this paper is made available at \url{https://doi.org/10.1007/s00354-021-00125-3}.

Please cite as: 

B. Paixão, L. Baroni, M. Pedroso, R. Salles, L. Escobar, C. de Sousa, R. de Freitas Saldanha, J. Soares, R. Coutinho, et al., 2021, Estimation of COVID-19 Under-Reporting in the Brazilian States Through SARI, New Generation Computing

\end{abstract}

\keywords{COVID-19 \and under-reporting \and time series \and SARI \and epidemiology}

\section{Introduction}

In January 2020, the new coronavirus (COVID-19) was considered a Public Health Emergency of International Importance by the World Health Organization (WHO). Later, in March, WHO characterized the disease as a pandemic. Due to its relevance, many efforts are being made to combat COVID-19, either by discovering the characteristics of the virus, methods of prevention, treatment, or directing public policy action \cite{callaway_coronavirus_2020}.

In Brazil, interventional measures such as the creation of field hospitals, surveillance information systems, and actions to reduce the economic impact are being adopted to mitigate the effects caused by COVID-19. Among the main objectives is the one to slow down the spread of the virus to avoid overloading the health system. In this sense, policies to encourage prevention are adopted, such as, for example, the recommendation or imposition of physical isolation and quarantine \cite{zheng_risk_2020}.

Decision-making for the adoption of public policies in this pandemic scenario is stressing and, at the same time, challenging task. Part of the difficulty comes from the lack of specific information about essential characteristics such as the total number of people infected. There is a lack of availability of tests to confirm the infection by SARS-CoV-2, which ends up being performed only in more severe cases of the disease, with exceptions. Such a scenario makes the capacity of the health system to monitor the evolution of the number of cases uncertain. The discrepancy between the actual amount of infected and diagnosed individuals constitutes under-reporting \cite{marson_covid-19_2020}.

It is estimated that under-reporting is a relevant factor in determining the actual mortality rate and, if not considered, can cause significant misinformation \cite{lachmann_correcting_2020}. Therefore, the objective of this work is to estimate the under-reporting of cases and deaths of COVID-19 in Brazilian states. If the possibility of testing the entire population is not viable, data from the Infogripe on notification of Severe Acute Respiratory Infection (SARI) are used.

The estimate of real cases of the disease, called novelty, is calculated by comparing the difference in SARI cases in 2020 (after COVID-19) with the total expected cases in recent years (2016 to 2019) derived from a seasonal exponential moving average. The novelty is based on inertial concepts. That is, there is a strength to maintain the values of a time series in a stable state over time \cite{gujarati_basic_2002}. Inertia remains until a rupture occurs. In this case, the rupture is the influence of the COVID-19. Under-reporting, then, is given by the difference between the novelty and the number of reported cases. In the end, under-reporting (cases and deaths) is presented as a rate for each state in Brazil.

Our paper stands out for estimating the under-reporting of cases and deaths of COVID-19 in Brazilian states. The methodology adopted includes everything from data acquisition and pre-processing to the calculation of under-reporting rates. Event detection methods are used to determine the parameters to be used in the methodology, and the estimate considers the weighted historical record. It adds value to the analysis, allowing a view more faithful to reality.

The results show that under-reporting rates vary significantly between states and that there is no standard for states in the same region in Brazil. It is noticed that the rates of under-reporting of cases are higher in the states of Minas Gerais (MG) and Mato Grosso do Sul (MS), and the highest rate of under-reporting of deaths is in the state of MG. In addition to the under-reporting rates, a brief exploratory analysis is presented, showing some interesting investigations that may help to understand the initial process of the COVID-19 pandemic situation in the country, as well as to analyze epidemic moments in last years.

This article is divided into seven sections in addition to this introduction. In Section \ref{sec:background}, the theoretical foundation that supports the adopted methodology is presented, whereas Section \ref{sec:related-work} presents a summary of the published works regarding the under-reporting of COVID-19. Section \ref{sec:methodology} discusses the process of under-reporting estimation. Section \ref{sec:experiments} presents the experimental setup of the scenario in which the methodology was applied. Section \ref{sec:results} presents the most relevant search results. Finally, in Section \ref{sec:final_remarks} the main conclusions of the work are pointed out.

\section{Background} \label{sec:background}

In this section, we introduce some background for time series (Section \ref{subsec:time_series}), moving averages (Section \ref{subsec:medias_moveis}), and event detection (Section \ref{subsec:deteccao_eventos}) used in the context of this work.

\subsection{Time Series}
\label{subsec:time_series}

A time series is a sequence of observations collected in time. Usually, a time series $y$ can be considered as a stochastic process, \emph{i.e.}, a sequence of $n$ random variables \textless $y_1, y_2, \cdots, y_n$\textgreater \cite{esling_time-series_2012,shumway_time_2017}. A specific observation of a time series is represented as $y_i$, indexed in time by $i = 1, \dots, n$, where $y_1$ represents the first observation and $y_n$ is the most recent observation.

The $i$-th subsequence of size $p$ in a time series $y$, represented as $seq_{i, p} (y)$, is a continuous sequence of values \textless$y_{i-(p-1)}, y_{i-(p-2)},$ $\ldots$, $y_{i}$\textgreater, where $|seq_{i, p} (y)|$ = $p$ e $p \le i \le |y| $. The sequence contains $i$-th observation and its $p-1$ predecessors.

The $i$-th subsequence outdated seasonally in periodicity $s$ of size $p$ in a time series $y$, represented as $seq_{i,p}^s(y)$, is an ordered sequence of values \textless$y_{i-(p-1)\cdot s},$ $y_{i-(p-2)\cdot s},$ $\ldots$, $y_{i}$\textgreater, where $|seq_{i,p}^s(y)|$ = $p$ and $p \le i \le |y|$. The sequence contains $i$-th observation and its $p-1$ predecessors outdated seasonally.

\subsection{Seasonal Moving Averages}
\label{subsec:medias_moveis}

The $i$-th moving average $\overline{y}_{i,p}$ of $p$ terms in a time series $y$ is calculated by the average of $t_k$ observations in the sequence $seq_{i,p}(y)$, as shown in Equation \ref{eq:sma}. The $i$-th exponential moving average $\hat{y}_{i,p}$ of $p$ terms in a time series $y$ is calculated by the weighted average of $t_k$ observations in the sequence $seq_{i,p}(y)$ and the weights $\alpha_k$. The $\hat{y}_{i,p}$ is described in Equation \ref{eq:sema}, where there is more emphasis on the most recent observations.

\begin{equation}\label{eq:sma}
\overline{y}_{i,p} = \frac{\sum_{k=1}^p{t_k}}{p} ~|~ t_k \in seq_{i,p}(y), ~ p \le i \le |y|
\end{equation}

\begin{equation}\label{eq:sema}
\hat{y}_{i,p} = \frac{\sum_{k=1}^p{\alpha_k \cdot t_k}}{\sum_{k=1}^p{\alpha_k}} ~|~ t_k \in seq_{i,p}(y), \alpha_k = \left(1 - \frac{2}{p + 1}\right)^{p-k}, ~ p \le i \le |y|
\end{equation}

The $i$-th seasonal moving average $\overline{y}_{i,p}^s$ and the $i$-th seasonal exponential moving average $\hat{y}_{i,p}^s$ of $p$ terms in a time series $y$ are similarly calculated replacing the continuous sequence $seq_{i,p}(y)$ with the seasonal sequence $seq_{i,p}^s(y)$, respectively, in Equations \ref{eq:sma} and \ref{eq:sema}.

\subsection{Event Detection}
\label{subsec:deteccao_eventos}

Event detection methods include the discovery of anomaly and change points. Anomalies are observations that stand out because they do not appear to have been generated by the same process as the other observations in the time series \cite{kuchar_outlier_2017}. Change points characterize a transition between different states in a process that generates the time series data \cite{takeuchi_unifying_2006,ding_multiple_2017}.

There are several methods to address the detection of anomalies \cite{chandola_anomaly_2009,gupta_outlier_2014} and change points \cite{aminikhanghahi_survey_2017}. Among them, there are methods that consider the effects of inertia on time series data. As this work is based on inertial concepts \cite{gujarati_basic_2002}, this section presents two methods of this group.

\subsubsection{Anomaly by Adaptive Normalization}

Adaptive Normalization \cite{ogasawara_adaptive_2010} is used to detect anomalies. This technique uses inertia to address heteroscedastic non-stationary series. Given a time series $y$, the outlier removal process consists of three stages: (i) inertia calculation, (ii) noise calculation, and (iii) anomaly identification. In the inertia calculation, a moving average for the series $\overline{y}_{i,p}$ with $p$ terms is calculated, as described by Equation \ref{eq:sma}. The higher the value of $p$, the greater the inertia and the lower the adaptation speed. The noise $\epsilon_i$ is calculated by the difference between $y_i$ and $\overline{y}_{i,p}$, \emph{i.e.}, $\epsilon_i = y_i - \overline{y}_{i,p}$. Finally, the observations $\epsilon_i$ classified as outliers by boxplot correspond to anomalies in Equation \ref{eq:boxplot}.

\begin{equation}\label{eq:boxplot}
anomaly(y) = \{i, \forall i ~|~ y_i \notin [Q_1(y) - 3 \cdot IQR(y), Q_3(y) + 3 \cdot IQR(y)]\}
\end{equation}

\subsubsection{Change Points by Change Finder}

Change Finder is a technique that detects change points in univariate time series data \cite{takeuchi_unifying_2006}. Given a time series $y$, the event detection process consists of two phases. In the first phase, outliers are detected. For this, a learning model $\xi$ is adjusted to the time series $y$, resulting in $\hat{y}_i = \xi(y)_i$\footnote{in this work, linear regression was used for adjustment.}. Next, a score $s_i$ is calculated for each observation in the series related to its deviation from the learned model. This calculation produces a time series $s$, as presented in Equation \ref{eq:score_outliers}. The highest scores for $s$, classified according to Equation \ref{eq:boxplot}, indicate the occurrence of anomalies.

In the second phase, change points are detected. For this, a new time series $\overline{s}_p$ is produced, composed of moving averages of $s$ with $p$ terms, according to Equation \ref{eq:sma}. The detection of change points is then reduced to the outlier detection problem in $\overline{s}_p$ like the first phase.

\begin{equation}\label{eq:score_outliers}
s_i = \left(\hat{y}_{i}-y_{i}\right)^2, \ \hat{y}_i = \xi(y)_i
\end{equation}

\section{Related Work} \label{sec:related-work}

Due to its relevance and novelty, COVID-19 has been attracting much interest in the academy. Therefore, many works on COVID-19 have been published since the beginning of 2020 until today. However, there are still few studies focused on under-reporting estimates.

Looking for similar work, we searched in the Scopus database in May 2020 with the search string ((``covid-19'' OR ``covid19'') AND (``sub-notification'' OR ``under-reporting'' OR ``under-reporting'')). Only four papers in English were returned by the search. This low number of related publications can be a consequence of the time spent on the execution, review, editing, and publication of papers in scientific journals. Therefore, we accomplished a search for academic works in Google Scholar to complement the research, employing the same words as the search string used in the Scopus database and on the same date.

From the returned works, ten were selected for reading. Most of them discuss the characteristics of COVID-19, such as under-reporting (cases and deaths) and its possible impact on different scenarios \cite{ricoca_peixoto_epidemic_2020,abbara_coronavirus_2020}. Some works address the specificities of COVID-19 together with other diseases and the under-reporting rate as a factor to be considered \cite{mohindra_radiation_2020,ong_covid-19_2020}. Others make different estimates related to COVID-19 and cite the under-reporting as a limitation or parameter \cite{lau_internationally_2020,russell_using_2020}. Three of the returned works are more specific regarding the under-reporting estimate, being more closely related to this work \cite{krantz_level_2020,lachmann_correcting_2020,ribeiro_estimate_2020}.

Krantz et al. \cite{krantz_level_2020} used harmonic analysis and wavelets to model the under-reporting of COVID-19 in several countries around the world. They developed susceptibility and infection equations with parameters varied according to the characteristics of each country to build adaptive models. The under-reporting rate was calculated by the difference between the numbers predicted by the model and reported numbers. The result provided the ratio between reported and unreported cases in the format ($1$ to $x$) in seven countries. The authors concluded that the results are not entirely accurate due to the lack of some important information that should be included in the model and were not available.

Similarly, to review the numbers of reported COVID-19 cases in several countries, Lachmann et al. \cite{lachmann_correcting_2020} also estimated expected cases. For this, the author used demographic data and fixed mortality rates of the countries as well as the paired comparison with the reference country (South Korea). It presented and discussed estimates of the number of people infected with COVID-19 considering a certain set of situations that must be true to justify the model.

Ribeiro et al. \cite{ribeiro_estimate_2020} used regression techniques on hospitalization data in Brazil with a type of acute respiratory syndrome as the cause. They analyzed the time evolution of hospitalizations for each month in the period between 2012 and 2019. They created a mathematical function that replicates the typical behavior of cases of hospitalization for SARI. This function was compared with data from 2020 in the same months to estimate under-reporting. The results showed an under-reporting rate of $7.7$:$1$ for Brazil.

Our work stands out for estimating the under-reporting of COVID-19 in Brazilian states weekly. In addition to under-reporting rates being calculated by week and by state, more detail than the cited works, the estimate considers the weighted historical record (in which most recent years have more weight than less recent ones) to predict expected SARI cases in 2020. It enriches the analysis allowing an estimate closer to reality. This work can also be highlighted for focusing on time series and using event detection tools in the study.

\section{Methodology} \label{sec:methodology}

In seasonal phenomena, time series are generated by superimposing a seasonal process and random noises. Based on this premise, Equation \ref{eq:pred} models the seasonal component of the time series, where $y_i$ is an observation, $\hat{y}_{i-s,p}^s$ is the seasonal exponential moving average (SEMA) in the previous seasonality and $\epsilon_{i}$ is the random noise. The obtained seasonal component brings up the inertia concept in time series. It enables the analysis of the intrinsic random noise of the observed phenomenon, while the influences that determine the behavior of the series are not changed \cite{gujarati_basic_2002}.

\begin{equation} \label{eq:pred}
y_i - \hat{y}_{i-s,p}^{s} + \epsilon_{i} = 0
\end{equation}

In the case of rupture (\emph{i.e.}, a ``break'' in inertial behavior), we adopt the concept of novelty $\eta$. The novelty is the influence introduced in each interval resulting from a rupture in a time series. Once the novelty begins, the modeled SEMA from past data is no longer the only representative process of the new behavior of the time series. In this context, Equation \ref{eq:pred} is expanded to Equation \ref{eq:novelty}, that expresses novelty $\eta_{i}$ and error $\hat{\epsilon}_{i}$. We have that $\hat{\epsilon}_{i}$ is approximated by the average error $\overline{\epsilon}$ observed in the pre-novelty period, \emph{i.e.}, $\hat{\epsilon}_{i}$ is expected to be inside the interval confidence for $\overline{\epsilon}$ ($[\overline\epsilon_{min}-\overline\epsilon_{max}]$).

\begin{equation} \label{eq:novelty}
y_i - \hat{y}_{i-s,p}^{s} - \eta_{i} - \hat{\epsilon}_{i} = 0, ~ \hat{\epsilon}_{i} \approx \overline{\epsilon}, ~ \hat{\epsilon}_{i} \in [\overline\epsilon_{min}-\overline\epsilon_{max}]
\end{equation}

Until the seasonal component $\hat{y}_{i-s,p}^{s}$ incorporates the novelty $\eta_{i}$, $\eta_{i}$ defines a new phenomenon in the time series. Regarding SARI, we assume that $\eta_{i}$ is directly associated with COVID-19, \emph{i.e.}, the new known phenomenon.

From this concept, we first compute the inertial behavior of the time series to estimate under-reporting. Let $t$ be the period in which the rupture $y_t$ occurs. In novelty period (\emph{i.e.}, $t \le i \le |y|$), $\eta_i$ is the subtraction of the observations of the time series $y_i$ by the values of SEMA from the previous period $\hat{y}_{i-s,p}^{s}$ and the error $\hat{\epsilon}_{i}$ (approximated by $\overline{\epsilon}$). Equation \ref{eq:novelty} shows the calculation of the time series with $\eta_i$ for each $i$ in the novelty period. The novelty $\eta_i$ estimates the brute number of observations that exceed the expected according to the inertial behavior of the time series and its fundamental error.

To estimate the brute number of under-reported time series, we use the number of observations classified as SARS-CoV-2 (Severe Acute Respiratory Infection Coronavirus 2) in the novelty period. Equation \ref{eq:subn} presents the calculation of the time series with absolute numbers of under-reported observations, where $cov_{i}$ are observations classified as SARS-CoV-2.

\begin{equation} \label{eq:subn}
cur_{i} = \sum_{i=t}^{|y|} \eta_{i} - \sum_{i=t}^{|y|} cov_{i}, ~ t \le i \le |y|
\end{equation}

As we assume that the modeled novelty in time series $\eta_{i}$ represents COVID-19 cases, the time series $sub_{i}$ defines the number of under-reported observations per week. Then, the estimates $sub_{i}$ are added together to form the accumulated number of under-reported observations in the period, represented as $cur_i$ in Equation \ref{eq:sum}.

\begin{equation} \label{eq:sum}
cur_i = \sum_{i=t}^{|y|} sub_{i}, ~ t \le i \le |y|
\end{equation}

The under-reporting rate is estimated by dividing the accumulated number of under-reported time series $cur_i$ by the accumulated number of total time series $cov_{i}$ for the period. Equation \ref{eq:taxa} describes the under-reporting rate, denoted as $tx_i$, where $tx_{|y|}$ is the final rate. In this work, this calculation provides the estimated under-reporting rates for cases and deaths of COVID-19 for each Brazilian state individually. Thus, these rates allow for a comparable interpretation between the states.

\begin{equation} \label{eq:taxa}
tx_i = \frac{cur_i}{cov_{i}}
\end{equation}

\section{Experimental Setup} \label{sec:experiments}

This section discusses the experimental setup of the scenario in which the methodology was applied. Section \ref{subsec:dados} presents the process of data acquisition and preparation, whereas Section \ref{subsec:selecao} describes the methods and parameters applied in the analysis. Section \ref{subsec:implementation} presents the implementation details.

\subsection{Data Acquisition and Preparation} \label{subsec:dados}

InfoGripe is the primary dataset used for the analysis and development of the work\footnote{Data collected on May 28$^{th}$, 2020}. It is an initiative of the Oswaldo Cruz Foundation (Fiocruz) with the Getulio Vargas Foundation (FGV) and the Brazilian Health Surveillance System of the Ministry of Health. It records weekly SARI reported cases since January 2009. These data come from the Influenza Epidemiological Surveillance Information System (SIVEP-Influenza) and present the cases following the criteria: (fever) AND (cough OR sore throat) AND (dyspnoea OR oxygen saturation $<$ 95\% OR respiratory difficulty) AND (hospitalization OR death), symptoms equivalent to SARI international records \cite{infogripe_boletim_2020}. 
For the sake of simplicity, we are calling the dataset $DT\_SARI$.

To keep only the relevant data, we apply the following filter: $type =$ ``State'' $\wedge$ $gender =$ ``Total'' $\wedge$ $scale =$ ``Cases''. The resulting dataset shows the number of cases or deaths per epidemiological week of a given year for each state. Besides, it specifies the number of observations that correspond to Influenza A, Influenza B, SARS-CoV-2, Respiratory Syncytial Virus (RSV), Parainfluenza 1, Parainfluenza 2, Parainfluenza 3, and Adenovirus.

It is then performed the differentiation of the case observations that evolved to death. For this, we apply a second filter that resulted in two datasets, one with cases ($DT\_SARI\_c$) and another with deaths ($DT\_SARI\_d$). Finally, five attributes of interest are selected: \field{Year}, \field{Week}, \field{State}, \field{Total}, and \field{SARS-CoV-2}. Table \ref {table:atributos} describes these attributes.

\begin{table}[!ht]
	\centering
	\caption{Attributes of processed datasets $DT\_SARI\_c$ and $DT\_SARI\_d$}
	\small{
		\begin{tabular}{L{2.5cm} L{9.8cm}}	
			\hline\noalign{\smallskip}
			Attribute & Description \\
			\hline\noalign{\smallskip}
			Year & the epidemiological year of first symptoms \\
			Week & the epidemiological week of first symptoms \\
			State & the state name \\
			Total & the total number of recorded cases ($DT\_SARI\_c$) / deaths ($DT\_SARI\_d$) \\
			SARS-CoV-2 & the total number of cases with positive results for COVID-19 ($DT\_SARI\_c$) / deaths by COVID-19 ($DT\_SARI\_d$) \\
			\hline\noalign{\smallskip}
		\end{tabular}
	}
	\label{table:atributos}
\end{table}

In addition to these data, we use the number of confirmed cases ($DT\_HM\_c$) and confirmed deaths ($DT\_HM\_d$) from COVID-19 by state, provided by the Ministry of Health\footnote{Data collected on May 31$^{th}$, 2020.}. These numbers are updated daily on the COVID-19 Portal, the official communication channel on the epidemiological situation of COVID-19 in Brazil \cite{saude_secretaria_2020}. The values are used for purposes of comparison with the results obtained in this work.

\subsection{Method and Parameter Selection} \label{subsec:selecao}

The method and parameter selection are a determining factor for the quality of the results obtained in the research. This section aims at justifying the applied methodology, which includes the choice of the used dataset, and the methods and parameters adopted in the data analysis.

\paragraph{Datasets}

The most severe cases of COVID-19 manifest respiratory symptoms, such as difficulty in breathing or shortness of breath, and chest pain or pressure \cite{rothan_epidemiology_2020}, symptoms also present in Acute Respiratory Infection (ARI). Fever is another common symptom, even in mild cases of the disease. It is the reason for choosing of SARI data ($DT\_SARI$) instead of ARI data ($DT\_ARI$). $DT\_SARI$ is a subset of $DT\_ARI$. They differ only in the manifestation of fever. Therefore, we consider that the probable cases of COVID-19 with severe symptoms also present fever, making $DT\_SARI$ the most suitable dataset to estimate the under-reporting of the disease \cite{ksiazek_novel_2003,rota_characterization_2003}.

\paragraph{SEMA for Inertial Model}

It is necessary to identify the SARI observations that correspond to the COVID-19 to compute the under-reporting of COVID-19 in Brazil. For this, data from years predating COVID-19 should be observed to model the expected inertial behavior if there was no pandemic. Thus, it is possible to estimate the COVID-19 case number as being the value that exceeds the expected for the same period in the year.

SEMA provides an appropriate method to create the inertial function since it is a trend indicator that assigns more weight to the most recent data considering a seasonal pattern. It is efficient to estimate the inertial behavior of a time series if the series has not undergone any significant behavior change in the period.

First, we define the time series for which SEMA is to be calculated. For this, three parameters are required: $p$, $i$, and $s$ (See Section \ref{sec:background}), where $i$ represents the time index of the reference time series, $p$ is the number of predecessors, and $s$ is the seasonality to be considered. Note that $p$ and $s$ are defined based on the locality of $i$.

The $s$ is chosen based on the seasonal variation of respiratory viral diseases. The annual epidemics of the common cold and the flu affect the human population of temperate regions in the winter season \cite{dowell_seasonality_2004,tchidjou_seasonal_2010,moriyama_seasonality_2020,chew_seasonal_1998}. Therefore, $s$ is defined as 52, since 52 corresponds to the number of weeks in the year. In this way, we guarantee the analysis of comparable observation sequences in the SARI series.

The parameters $p$ and $i$ are based on the response of the event detection algorithms. The event detection (targeting both change points and anomalies) in the series $DT\_SARI\_c$ and $DT\_SARI\_d$ consistently evidence, in several states, behavior change in two periods: \emph{(i)} between the end of 2015 and the beginning of 2016, and \emph{(ii)} between March and April 2020. Table \ref{table:dates} shows the dates of events detected in 2020 for each state.

\begin{table}[!ht]
	\centering
	\caption{Change point (CP) dates that occurred in 2020}
	\small{
		\begin{tabular}{C{0.75cm} C{1.9cm} C{1.9cm} || C{0.75cm} C{1.9cm} C{1.9cm}}
			\hline\noalign{\smallskip}
			UF	&	CP Cases	&	CP Deaths & UF	&	CP Cases	& CP Deaths \\
			\hline\noalign{\smallskip}
			AC	&	- 	&	-	 &	PB &	14/03/2020 	&	14/03/2020	\\
			AL	&	04/04/2020	&	04/04/2020	&	PE	&	21/03/2020	&	28/03/2020	\\
			AM	&	28/03/2020	&	04/04/2020	&	PI	&	14/03/2020	&	14/03/2020	\\
			AP	&	21/03/2020	&	28/03/2020	&	PR	&	- 	&	14/03/2020	\\
			BA	&	14/03/2020	&	21/03/2020	&	RJ	&	21/03/2020	&	28/03/2020	\\
			CE	&	28/03/2020	&	28/03/2020	&	RN	&	21/03/2020	&	14/03/2020	\\
			DF	&	14/03/2020	&	14/03/2020	&	RO	&	28/03/2020	&	28/03/2020	\\
			ES	&	14/03/2020	&	21/03/2020	&	RR	&	14/03/2020	&	14/03/2020	\\
			GO	&	14/03/2020	&	14/03/2020	&	RS	&	21/03/2020	&	21/03/2020	\\
			MA	&	22/02/2020	&	29/02/2020	&	SC	&	28/03/2020	&	14/03/2020	\\
			MG	&	14/03/2020	&	14/03/2020	&	SE	&	14/03/2020	&	14/03/2020	\\
			MS	&	14/03/2020	&	14/03/2020	&	SP	&	14/03/2020	&	14/03/2020	\\
			MT	&	14/03/2020	&	21/03/2020	&	TO	&	14/03/2020	&	18/04/2020	\\
			PA	&	04/04/2020	&	04/04/2020	&		& & 	 	\\
			\hline\noalign{\smallskip}
		\end{tabular}
	}
	\label{table:dates}
\end{table}

The events detected in 2020 are a consequence of COVID-19 in Brazil. These events coincide with the first record of the disease in the country, considering the time for the disease spread and the manifestation of symptoms \cite{bastos_covid-19_2020,ministerio_da_saude_boletim_2020}. The events appear from the 11th epidemiological week of 2020 for most states, \emph{i.e.}, two weeks after the first confirmed case of COVID-19 in Brazil (this occurred in the 9th epidemiological week of 2020).

This result identifies the beginning of the novelty period in the data ($t$), \emph{i.e.}, the 11th epidemiological week of 2020. Concerning the total number of weeks of the data, it corresponds to week 584 ($t=584$). So, the model should be executed for the period before this date and extended until the last week of data, which is the week 590 ($|y|$). The parameter $i$ admits values of the COVID-19 influence range (\emph{i.e.}, $584 \le i \le 590$). 

\begin{figure}[!ht]
	\centering
	\includegraphics[width=0.7\textwidth]{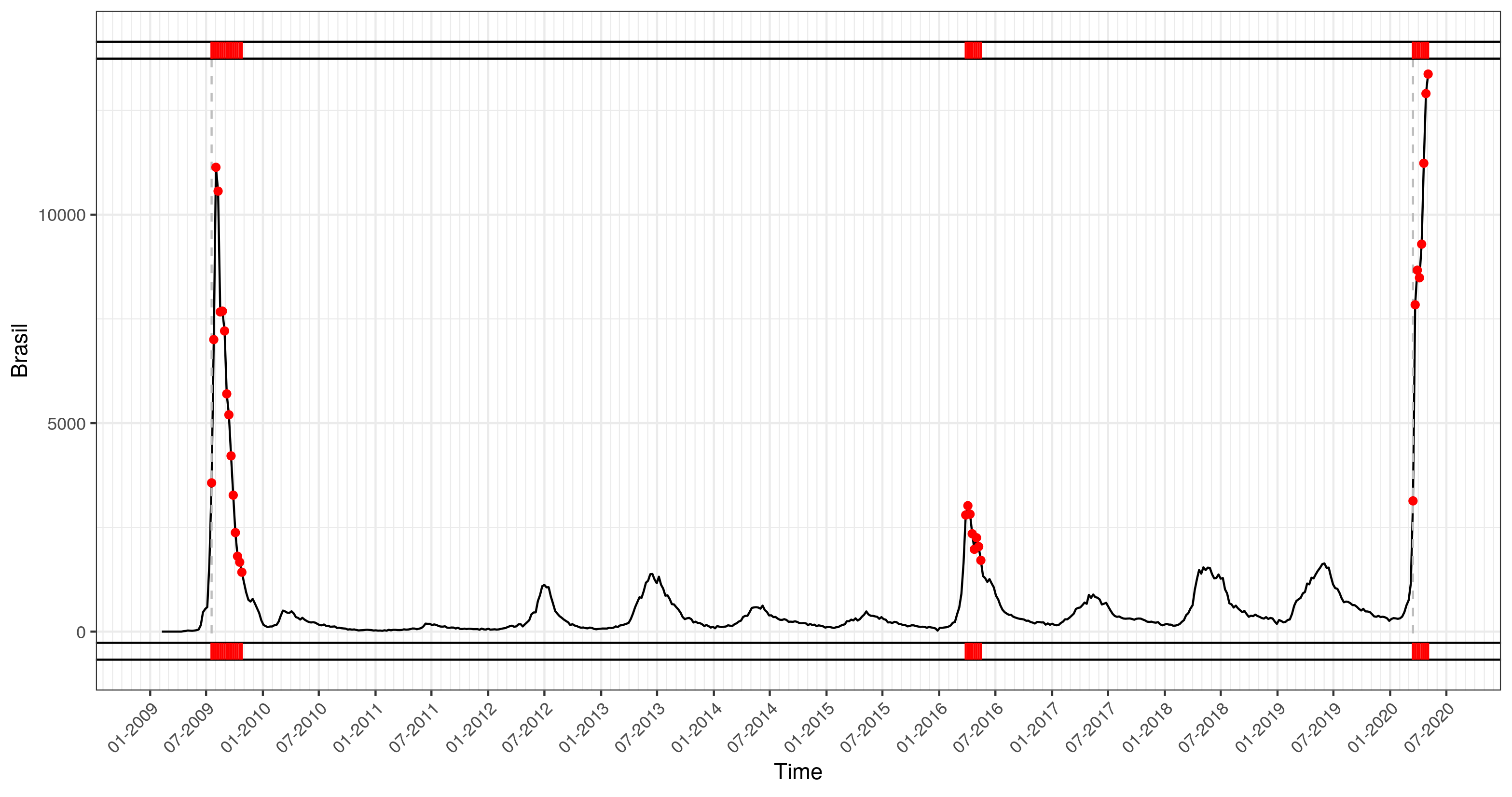}
	\caption{Events detected in the SARI cases curve in Brazil. The red dots mark anomalies (Adaptive Normalization), and the gray dotted lines mark the change points (Change Finder).}
	\label{fig:evt-brazil}
\end{figure}

Figure \ref{fig:evt-brazil} shows the events detected in the SARI cases curve in Brazil. In addition to 2009 (H1N1) and 2020 (COVID-19), events are observed in the 2015/2016 period. Events presented on this Figure correspond to abnormal behavior. They can affect the previous inertial behavior of the series. For this reason, the value attributed to $p$ is 4, meaning that the previous four years (2016 to 2019) are considered. 

Table \ref{table:parameters} summarizes the used parameters. The model errors (random noise) for this period for both the cases and deaths in each state are, respectively, described in Tables \ref{table:errors_c} and \ref{table:errors_d}. Since $\epsilon_i$ follows a non-normal distribution, the interval confidence for $\overline\epsilon$ is computed by bootstrap with 1000 repetitions. Under-reporting rates were calculated for states where it was found that there were, in fact, novelty. Therefore, average error observed in the pre-novelty period ($\overline{\epsilon}$) was compared with the novelty ($\eta_i$) and assessed whether there is a relevant difference at a significance level of 0.05 using the Wilcoxon test.

\begin{table}[ht!]
	\centering
	\caption{Parameters}
	\small{
		\begin{tabular}{C{2cm} C{3cm}}
			\hline\noalign{\smallskip}
			Parameter & Value \\
			\hline\noalign{\smallskip}
			$i$	&	$t \le i \le |y|$ \\
			$t$ &	$584$ \\
			$p$	&	$4$	 \\
			$s$	&	$52$ \\
			\hline\noalign{\smallskip}
		\end{tabular}
	}
	\label{table:parameters}
\end{table}

\begin{table}[ht!]
	\centering
	\caption{Errors of the models (cases)}
	\small{
		\begin{tabular}{C{0.75cm} R{1.1cm} R{2.0cm} || C{0.75cm} R{1.1cm} R{2.0cm}}
			\hline\noalign{\smallskip}
			UF & $\overline\epsilon$ & $[\overline\epsilon_{min}, \overline\epsilon_{max}]$ & UF & $\overline\epsilon$ & $[\overline\epsilon_{min}, \overline\epsilon_{max}]$ \\
			\hline\noalign{\smallskip}
			AC	&	1.727	&	[1.166, 2.344]	&	PB	&	2.198	&	[1.700, 2.821]	\\
			AL	&	1.482	&	[0.959, 2.092]	&	PE	&	11.537	&	[9.311, 13.81]	\\
			AM	&	9.770	&	[6.82, 14.343]	&	PI	&	2.651	&	[1.758, 3.979]	\\
			AP	&	0.299	&	[0.181, 0.478]	&	PR	&	24.465	&	[18.79, 31.21]	\\
			BA	&	10.211	&	[7.478, 13.31]	&	RJ	&	9.788	&	[6.514, 14.28]	\\
			CE	&	6.967	&	[4.372, 11.14]	&	RN	&	1.230	&	[0.705, 1.841]	\\
			DF	&	13.036	&	[11.19, 15.11]	&	RO	&	0.502	&	[0.162, 0.970]	\\
			ES	&	4.021	&	[2.789, 5.562]	&	RR	& -0.012	&	[-0.12, 0.119]	\\
			GO	&	6.349	&	[3.787, 10.31]	&	RS	& 7.516 	&	[1.965, 14.86]	\\
			MA	&	0.980	&	[0.617, 1.535]	&	SC	&	4.396	&	[1.316, 8.088]	\\
			MG	&	6.320	&	[1.449, 12.34]	&	SE	&	1.851	&	[1.391, 2.382]	\\
			MS	&	9.276	&	[6.668, 13.15]	&	SP	&	49.934	&	[21.59, 85.15]	\\
			MT	&	1.516	&	[0.855, 2.333]	&	TO	&	1.172	&	[0.909, 1.484]	\\
			PA	&	6.403	&	[5.012, 8.195]	&		&			&		\\
			\hline\noalign{\smallskip}
		\end{tabular}
	}
	\label{table:errors_c}
\end{table}

\begin{table}[ht!]
	\centering
	\caption{Errors of the models (deaths)}
	\small{
		\begin{tabular}{C{0.75cm} R{1.1cm} R{2.0cm} || C{0.75cm} R{1.1cm} R{2.0cm}}
			\hline\noalign{\smallskip}
			UF & $\overline\epsilon$ & $[\overline\epsilon_{min},\overline\epsilon_{max}]$ & UF & $\overline\epsilon$ & $[\overline\epsilon_{min}, \overline\epsilon_{max}]$ \\
			\hline\noalign{\smallskip}
			AC	&	0.480	&	[0.298, 0.688]	&	PB	&	0.586	&	[0.383, 0.815]	\\
			AL	&	0.293	&	[0.151, 0.481]	&	PE	&	0.325	&	[0.120, 0.555]	\\
			AM	&	0.670	&	[0.399, 1.094]	&	PI	&	0.185	&	[0.015, 0.376]	\\
			AP	&	0.047	&	[0.007, 0.100]	&	PR &	3.015	&	[2.129, 4.137]	\\
			BA &	0.847	&	[0.566, 1.182]	&	RJ	&	1.066	&	[0.563, 1.662]	\\
			CE	&	0.670	&	[0.378, 1.082]	&	RN &	0.409	&	[0.236, 0.613]	\\
			DF	&	0.423	&	[0.266, 0.603]	&	RO	&	0.056	&	[-0.028, 0.165]	\\
			ES	&	0.381	&	[0.161, 0.655]	&	RR	&	0.009	&	[-0.017, 0.050]	\\
			GO	&	0.940	&	[0.462, 1.466]	&	RS	&	0.902	&	[0.089, 1.717]	\\
			MA	&	0.093	&	[0.028, 0.169]	&	SC	&	0.632	&	[0.283, 1.075]	\\
			MG	&	0.993	&	[0.088, 2.061]	&	SE &	0.119	&	[0.049, 0.196]	\\
			MS	&	0.976	&	[0.460, 1.658]	&	SP	&	3.941	&	[1.110, 8.098]	\\
			MT	&	0.246	&	[0.046, 0.443]	&	TO	&	0.302	&	[0.198, 0.418]	\\
			PA	&	0.449	&	[0.226, 0.719]	&		&		 & 				\\
			\hline\noalign{\smallskip}
		\end{tabular}
	}
	\label{table:errors_d}
\end{table}

\subsection{Implementation} \label{subsec:implementation}

The adopted methodology was implemented in R \cite{r_core_team_r:_2014}. The code description and Jupyter notebook also developed in R complements this work\footnote{available at \url{https://eic.cefet-rj.br/~dal/covid-19-under-report/}}. In it, it is possible to check the entire process on the calculation of the under-reporting rates and all numerical and graphical results. The graphics with the cases and deaths series from the $DT\_SARI$ and the marking of the detected events are presented in this notebook for all states. Also, the site contains graphics with the evolution of under-reported records over the weeks after COVID-19 for each state. There it is possible to see whether under-reported records increase, decrease or remain constant over time.

For the execution of the event detection methods, Adaptive Normalization and Change Finder, the Harbinger\footnote{Available at \url{https://eic.cefet-rj.br/~dal/harbinger/}.} framework was used for detecting events in time series. It receives the time series and parameters and returns the detected events. Thus, it was not necessary to implement these two techniques, just to invoke them from Harbinger. The parameters used are those defined in Section \ref{subsec:selecao}.

For each state, two time series were submitted to the process described in Section \ref{sec:methodology}, both from the InfoGripe dataset on hospitalizations for SARI ($DT\_SARI$). The first is the weekly series with information on the number of registered SARI cases in the state, and the second is the weekly series with information on the number of SARI deaths. Under-reporting rates were calculated for states where it was found that there were, in fact, under-reported notification. Therefore, the number of novelty calculated ($\eta_i$) was compared with the number classified as SARS-CoV-2 at Infogripe data ($cov_i$) and assessed whether there is a relevant difference at a significance level of 0.05 using the Wilcoxon test.

\section{Results} \label{sec:results}

This work focuses on estimating under-reporting rates for cases and deaths of COVID-19. In Section \ref{subsec:exp-analysis} an exploratory analysis is conducted. It contains discussions that are based on the results of event detection (change points and anomaly) over the SARI time series. These findings bring valuable information to help understand the disease scenario in the most affected states. Besides, they helped to evaluate the choice of the method and the confidence of the estimates. Then, the actual under-reporting rates are presented in Section \ref{subsec:subn-rates}. 

\subsection{Exploratory Data Analysis} \label{subsec:exp-analysis}

\begin{figure}
	\centering
	\subfloat[Amazonas cases]{
		\label{cases_AM}
		\includegraphics[width=0.42\textwidth]{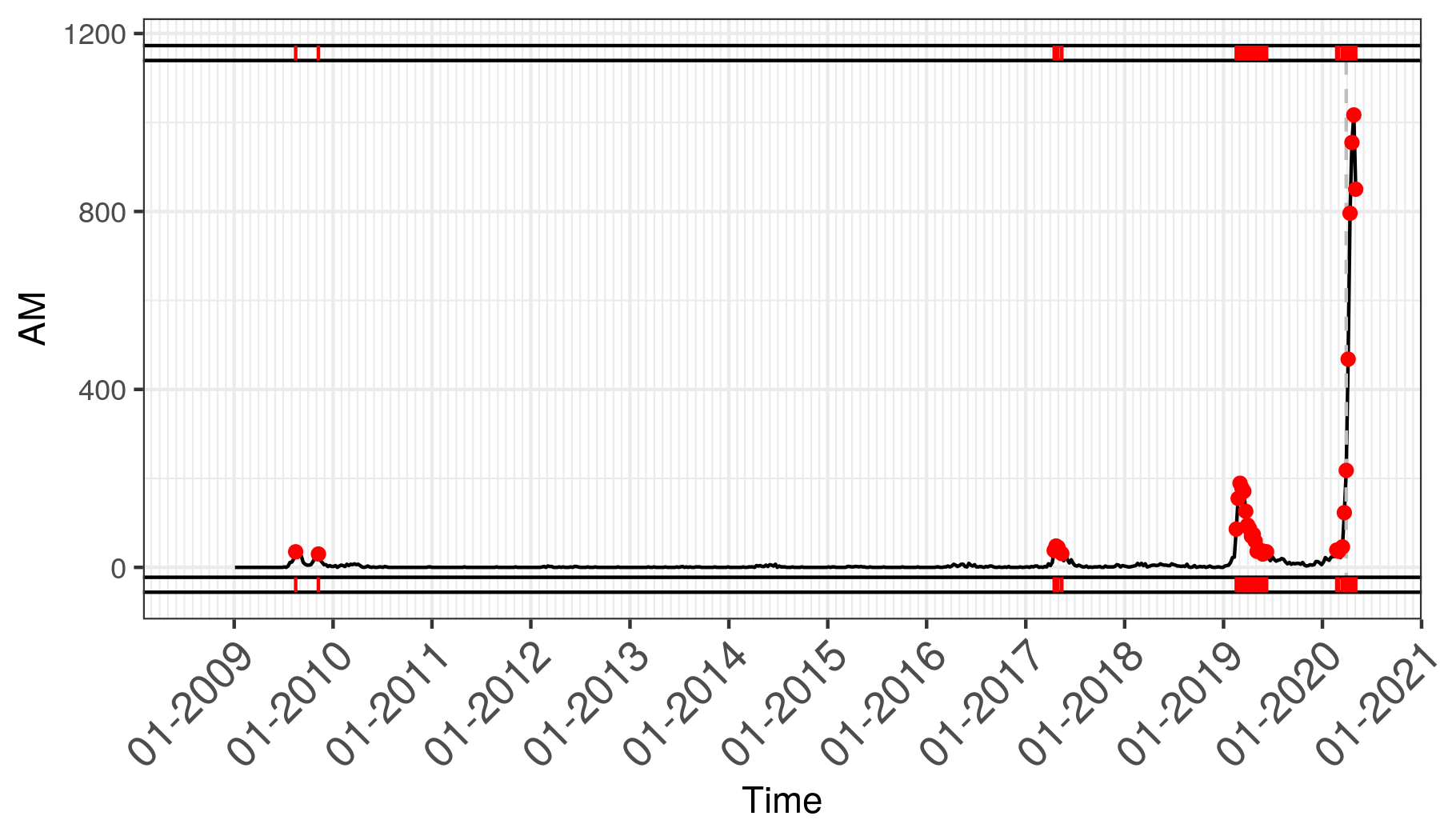}
	}
	\subfloat[Ceará cases]{
		\label{cases_CE}
		\includegraphics[width=0.42\textwidth]{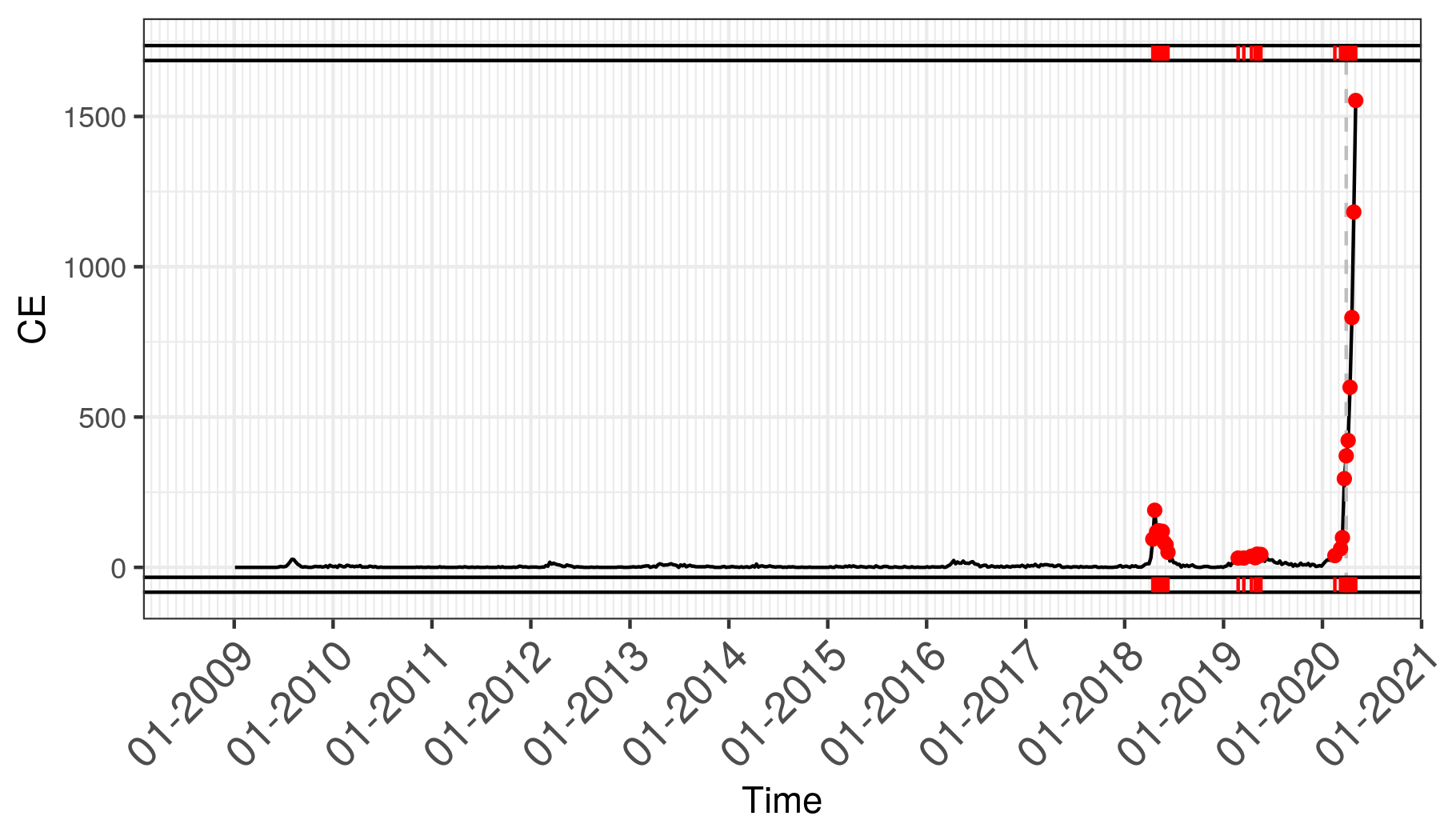}
	}
	
	\subfloat[Pernambuco cases]{
		\label{cases_PE}
		\includegraphics[width=0.42\textwidth]{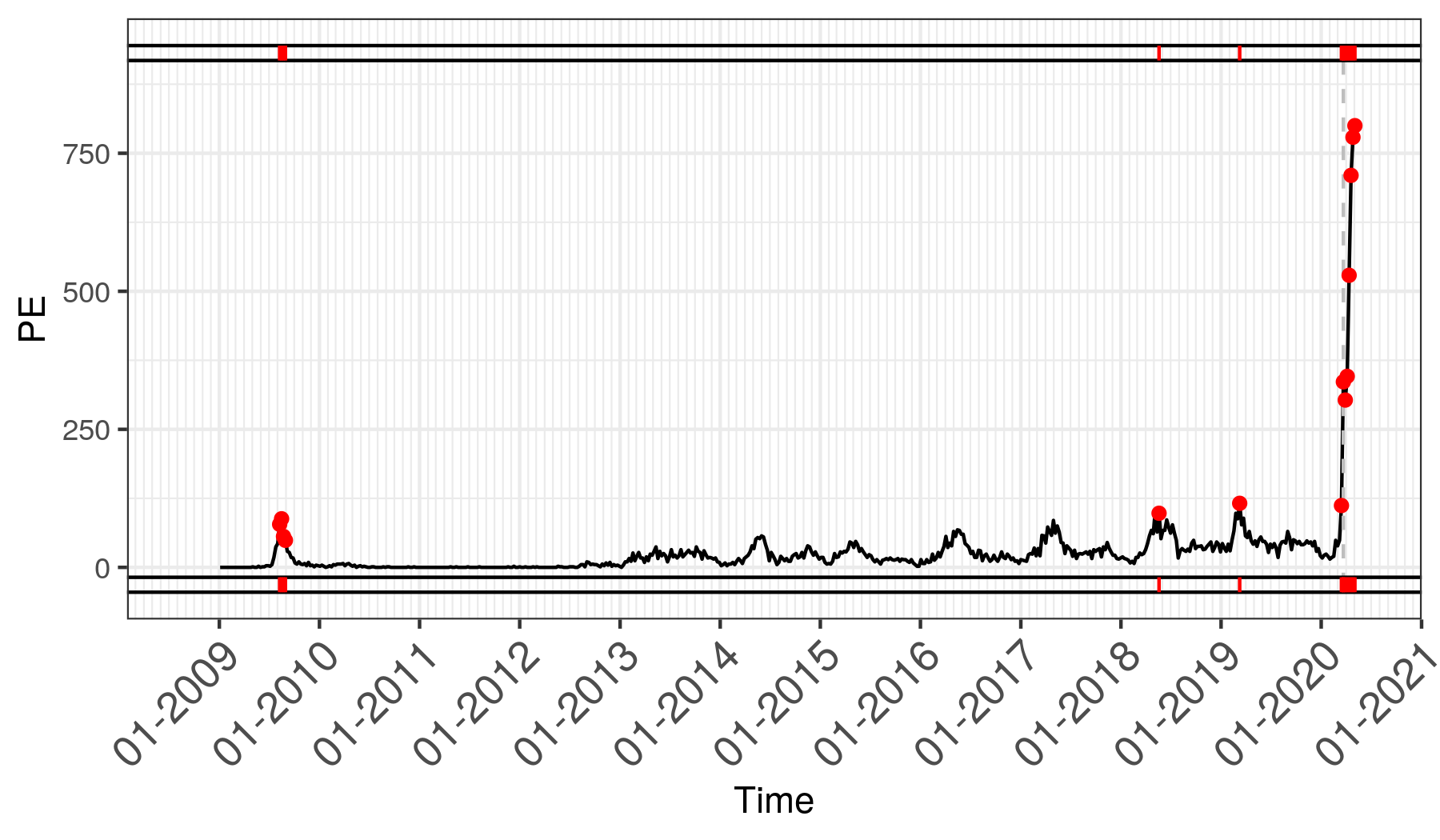}
	}
	\subfloat[Bahia cases]{
		\label{cases_BA}
		\includegraphics[width=0.42\textwidth]{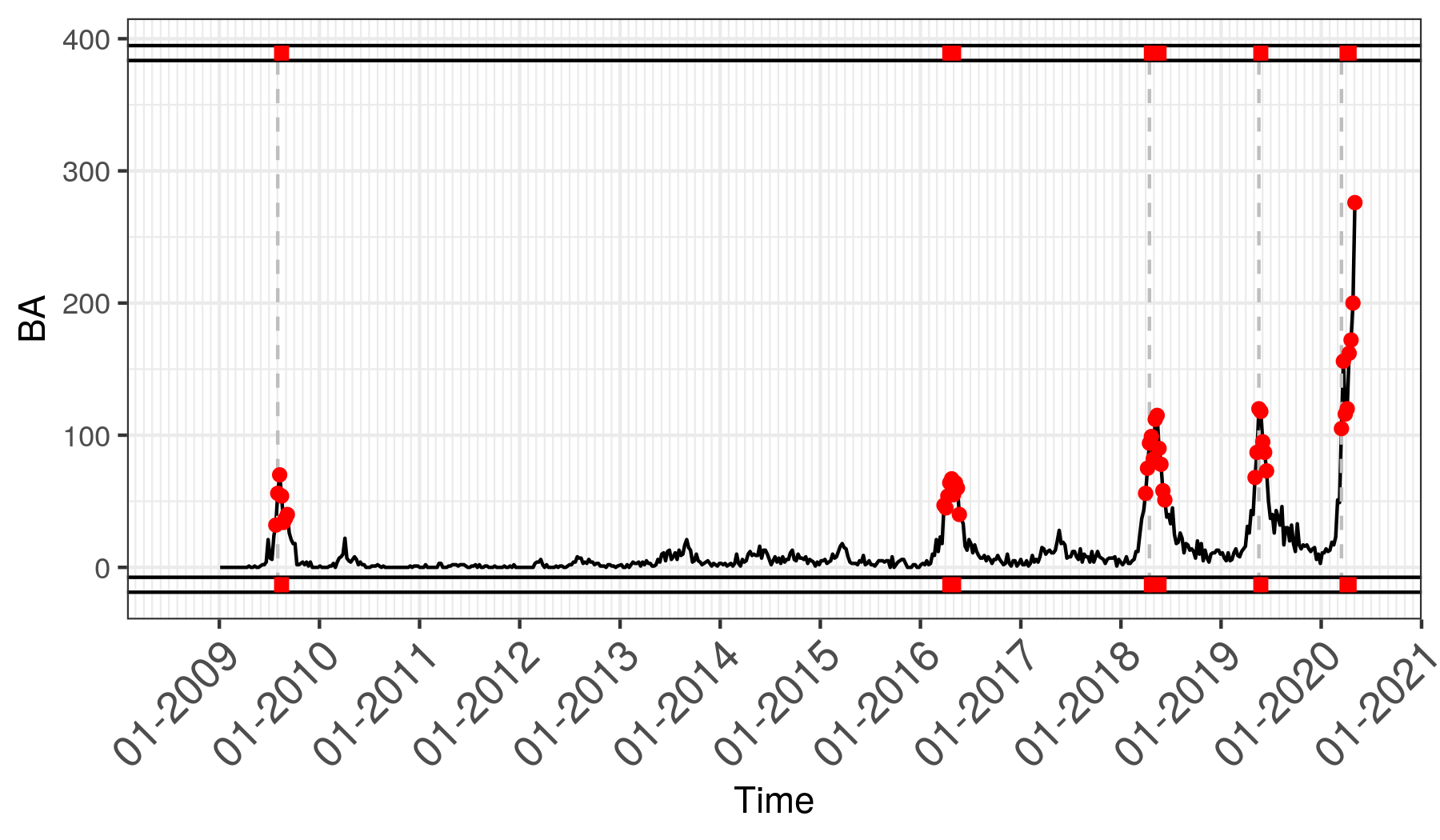}
	}
	
	\subfloat[Distrito Federal cases]{
		\label{cases_DF}
		\includegraphics[width=0.42\textwidth]{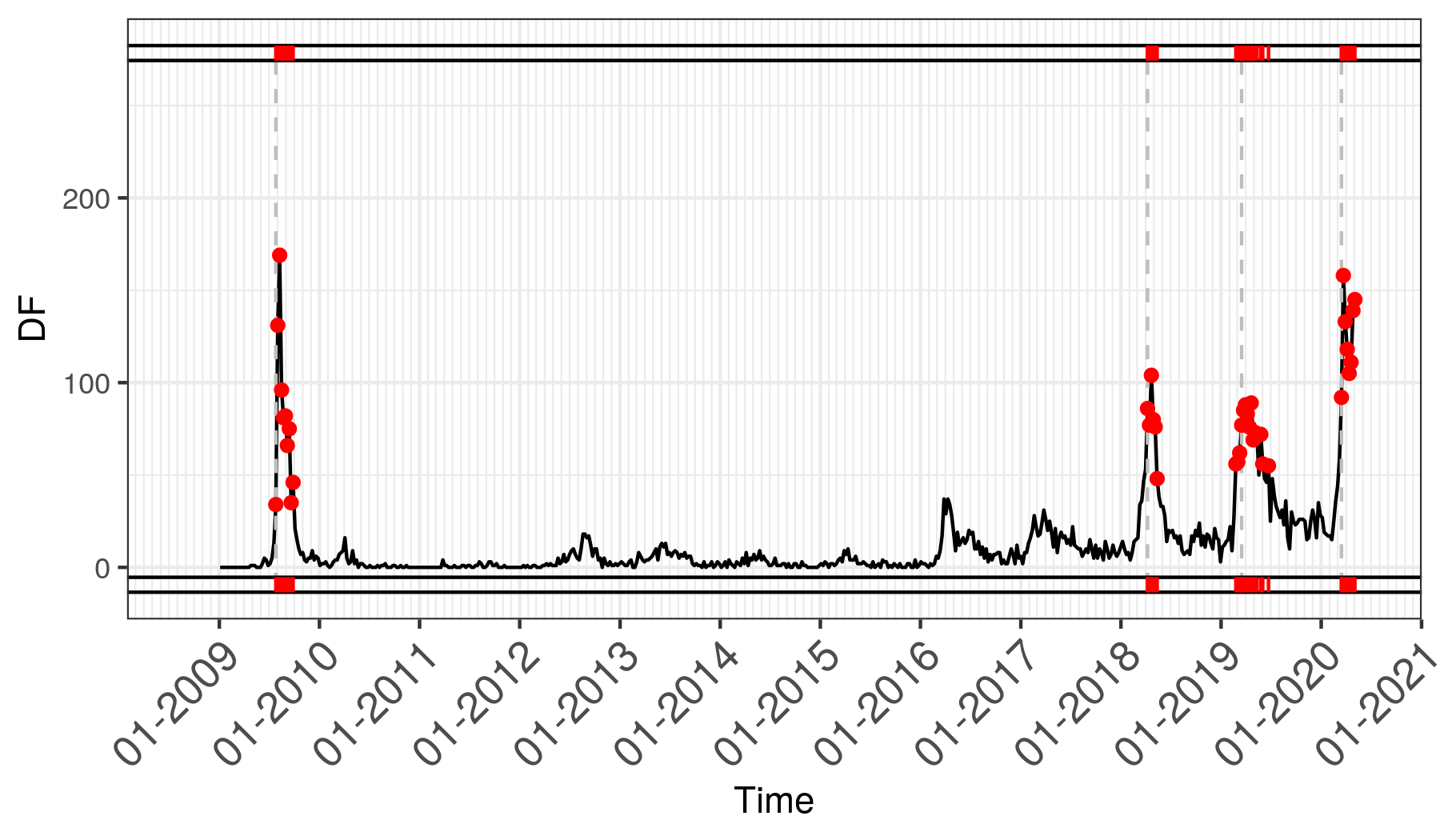}
	}
	\subfloat[São Paulo cases]{
		\label{cases_SP}
		\includegraphics[width=0.42\textwidth]{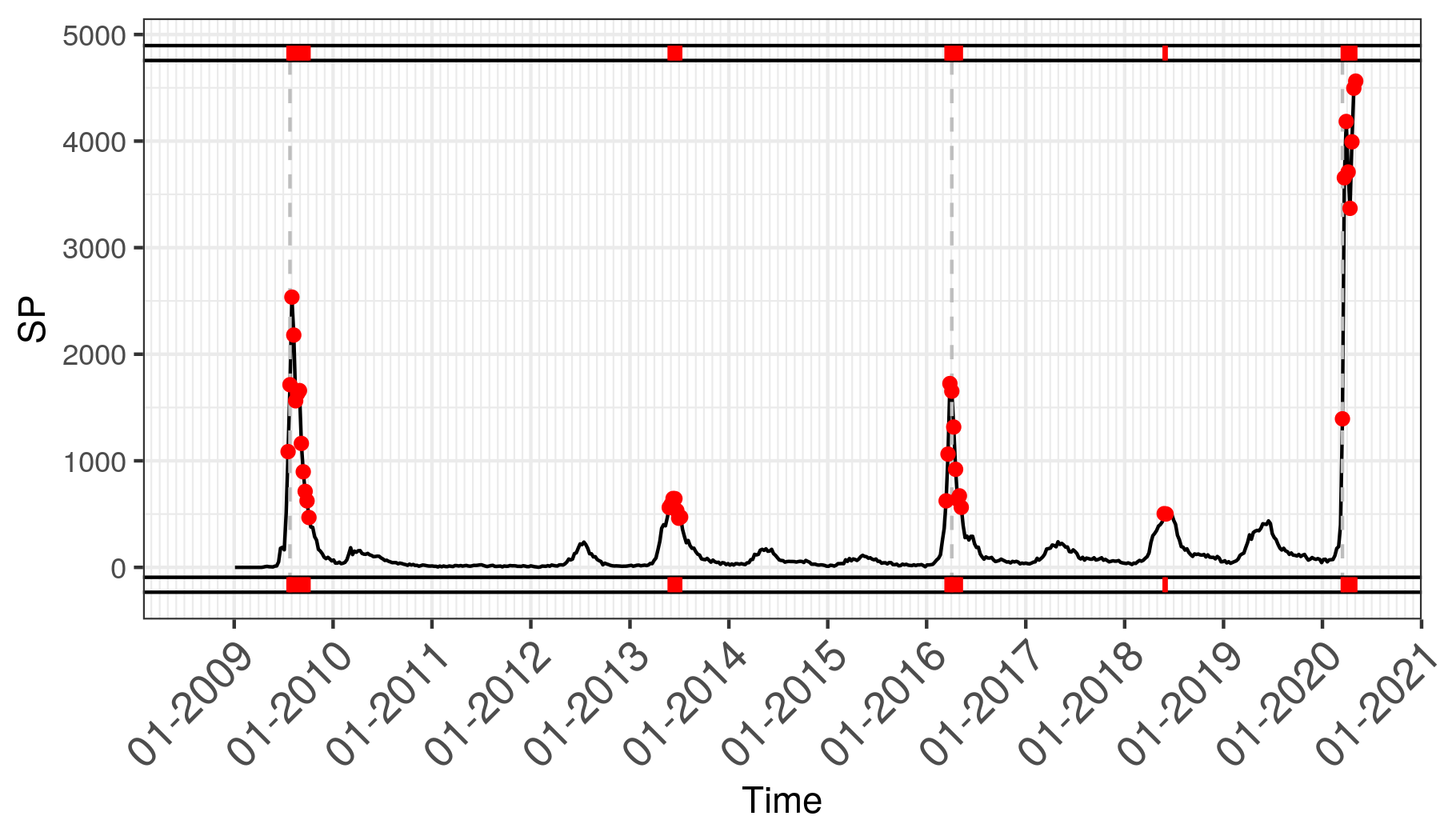}
	}
	
	\subfloat[Rio de Janeiro cases]{
		\label{cases_RJ}
		\includegraphics[width=0.42\textwidth]{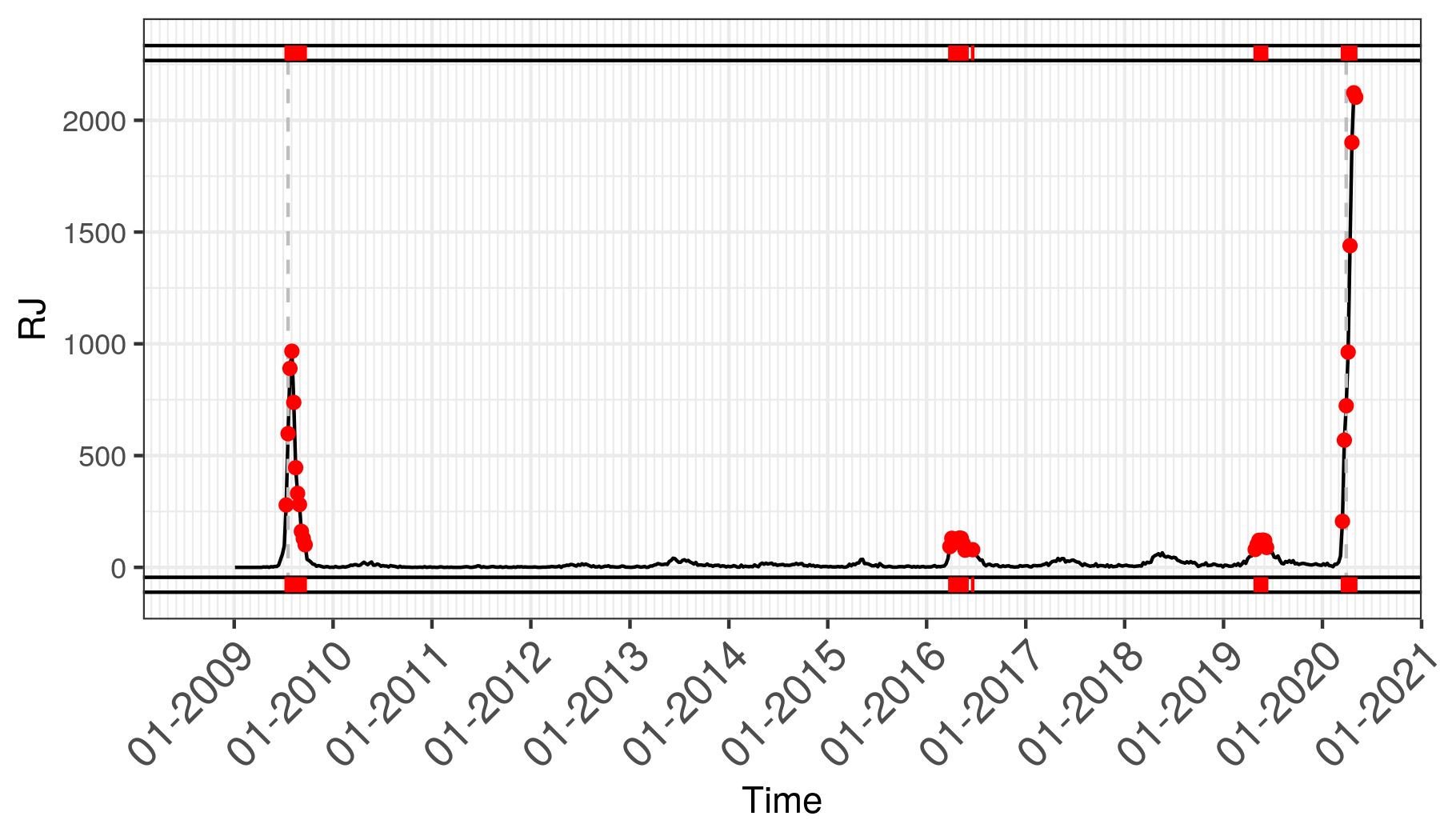}
	}
	\subfloat[Minas Gerais cases]{
		\label{cases_MG}
		\includegraphics[width=0.42\textwidth]{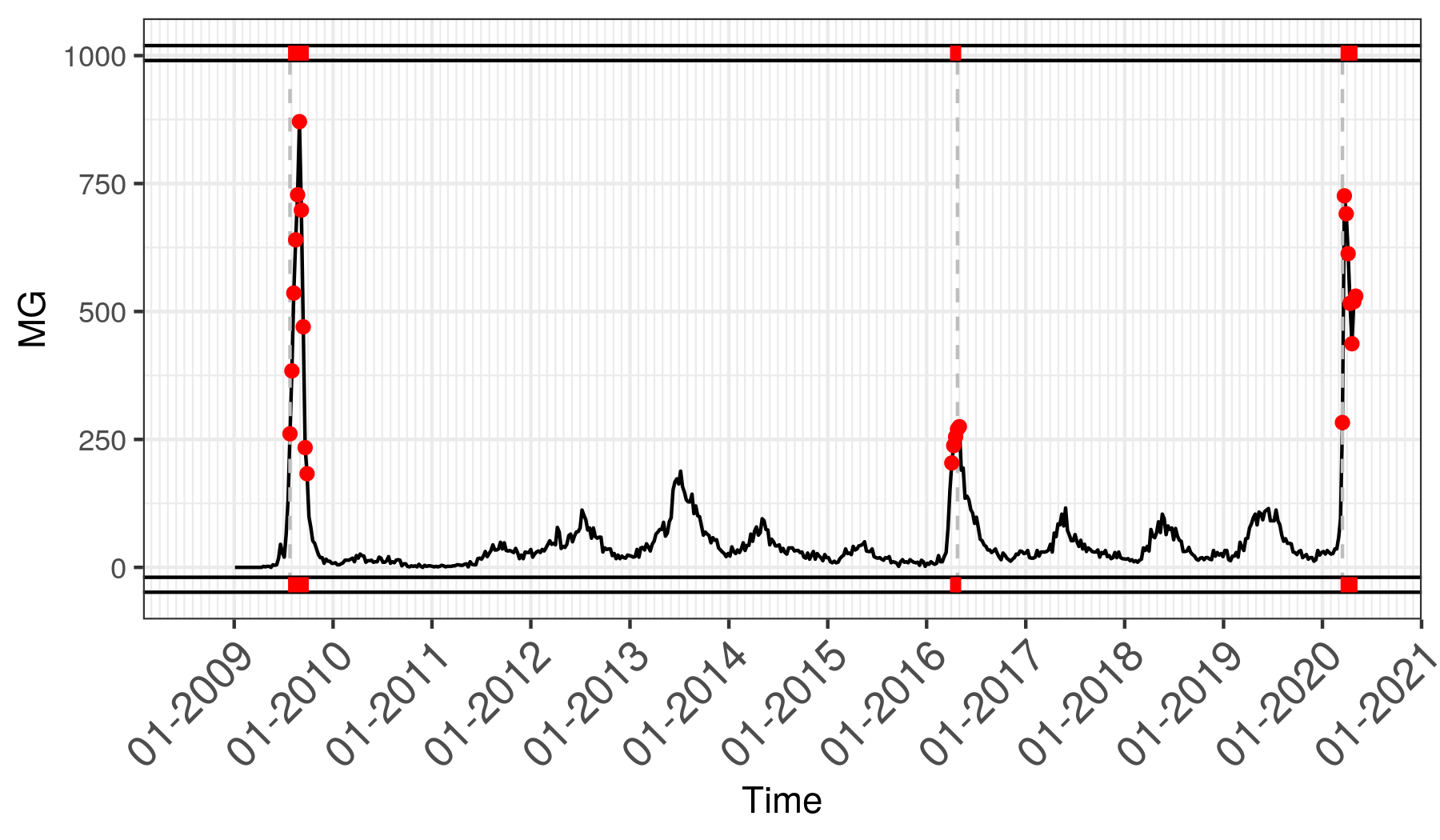}
	}
	
	\subfloat[Paraná cases]{
		\label{cases_PR}
		\includegraphics[width=0.42\textwidth]{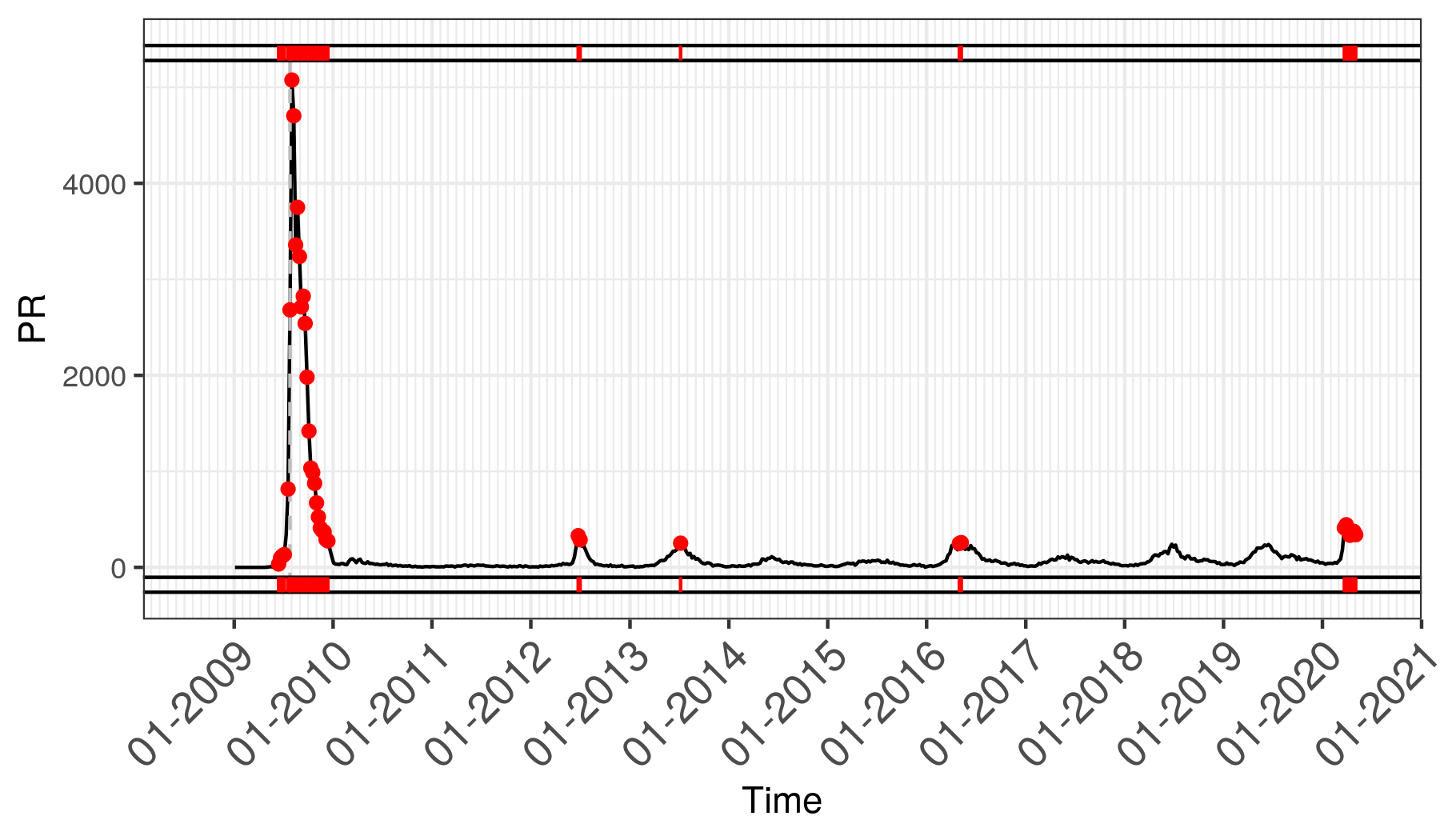}
	}
	\subfloat[Rio Grande do Sul cases]{
		\label{cases_RS}
		\includegraphics[width=0.42\textwidth]{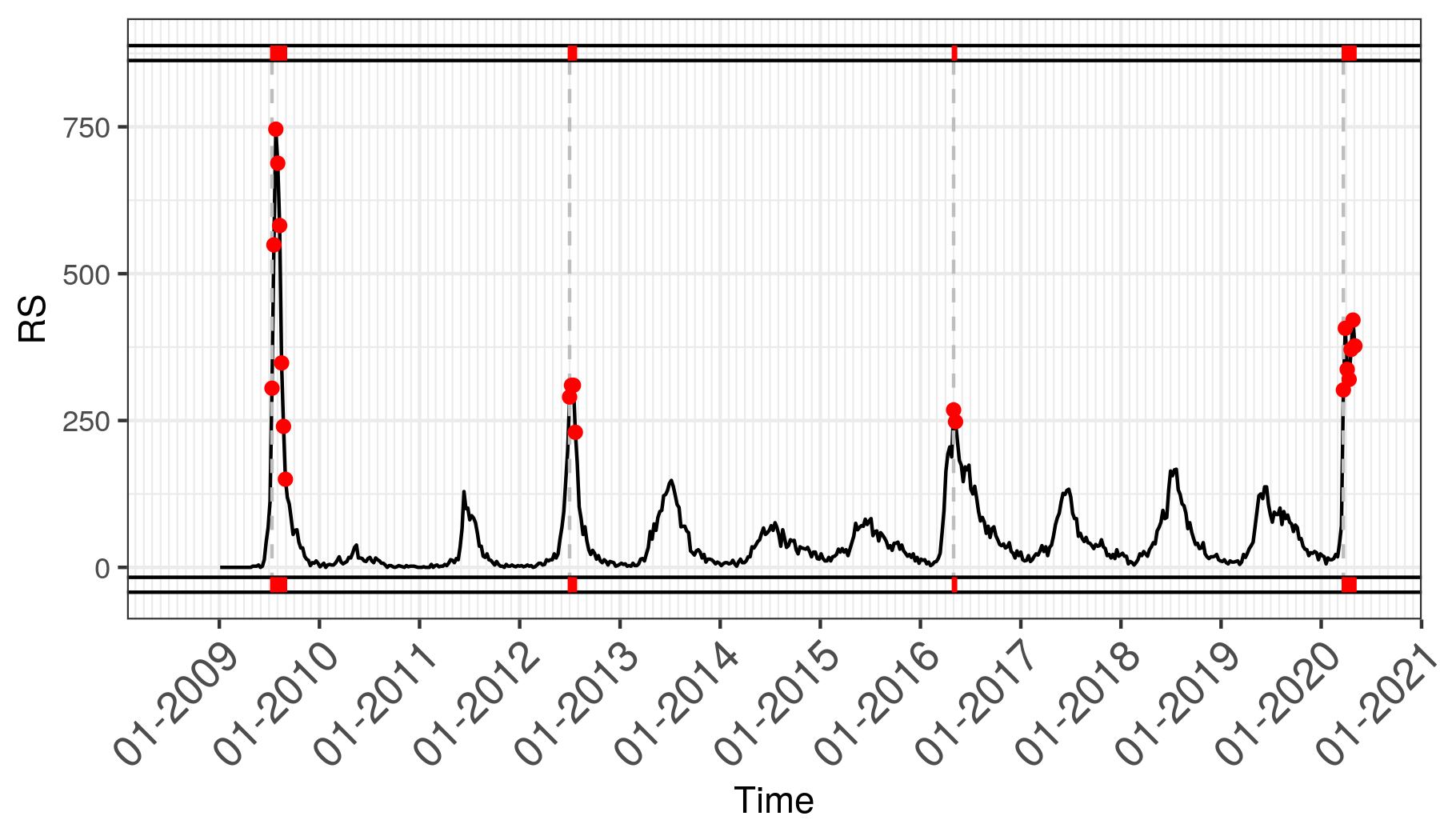}
	}
	\caption{Event detection in time series of cases}
	\label{fig_cases}
\end{figure}

\begin{figure}
	\centering
	\subfloat[Amazonas deaths]{
		\label{deaths_AM}
		\includegraphics[width=0.42\textwidth]{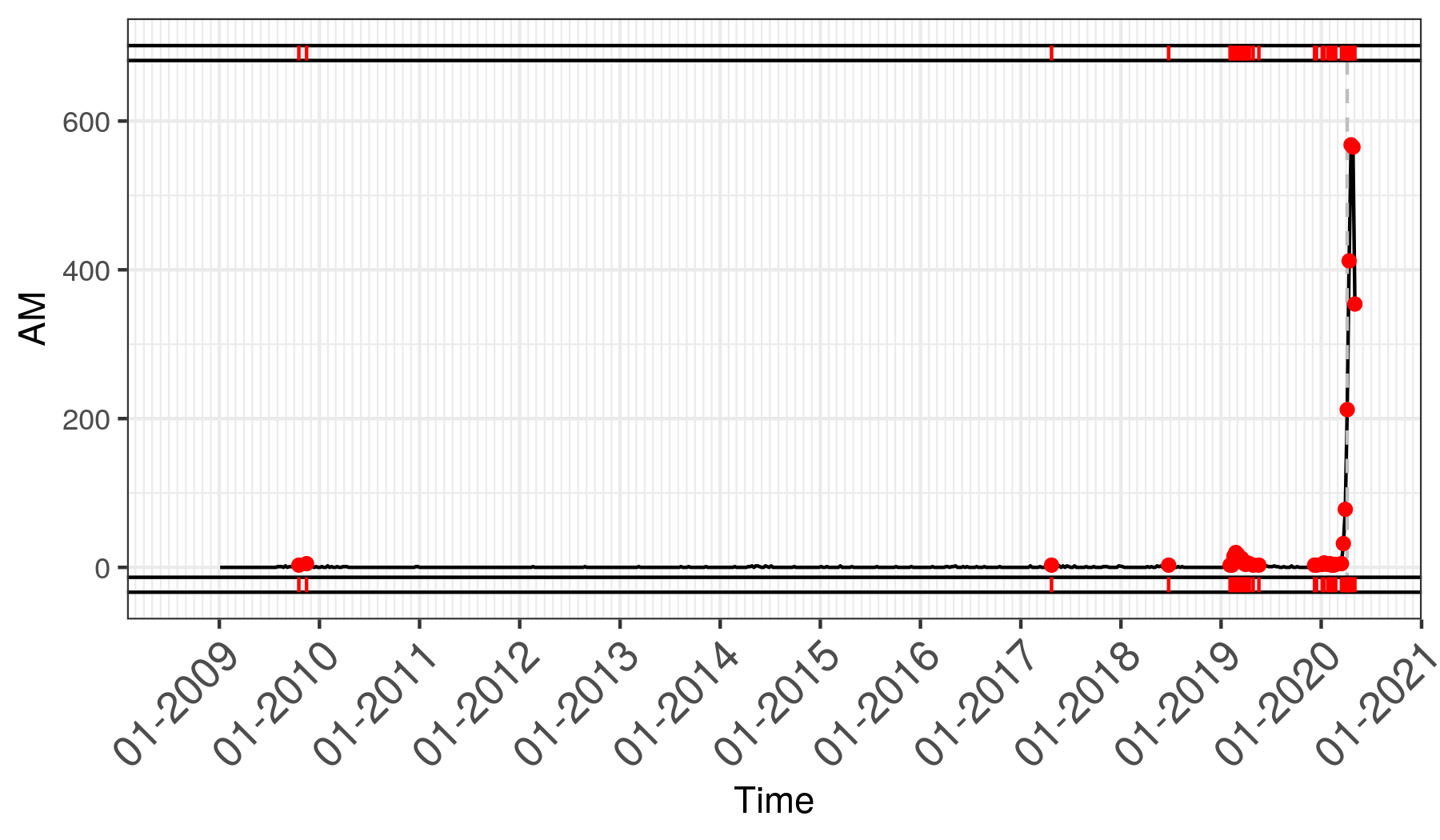}
	}
	\subfloat[Ceará deaths]{
		\label{deaths_CE}
		\includegraphics[width=0.42\textwidth]{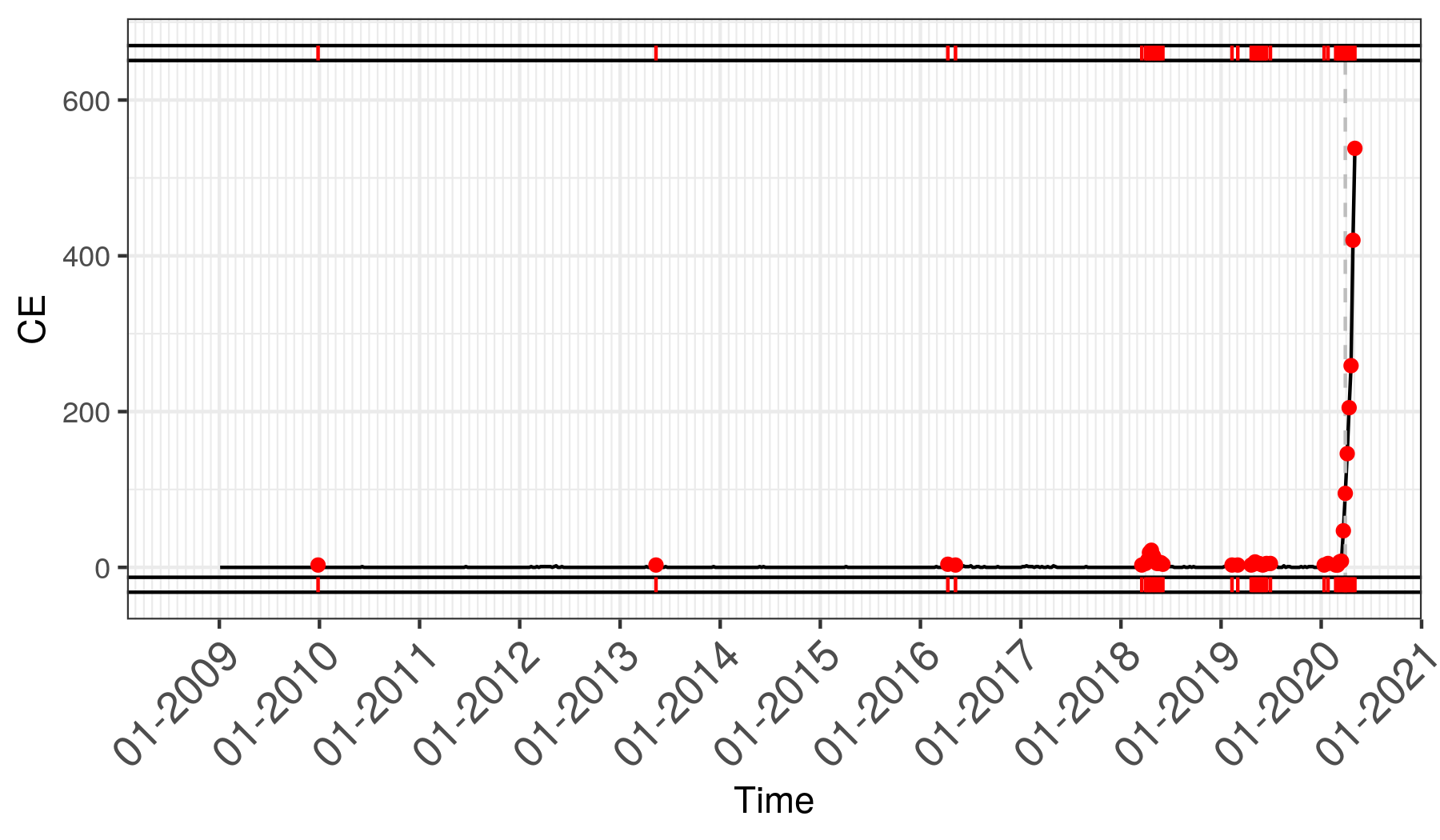}
	}
	
	\subfloat[Pernambuco deaths]{
		\label{deaths_PE}
		\includegraphics[width=0.42\textwidth]{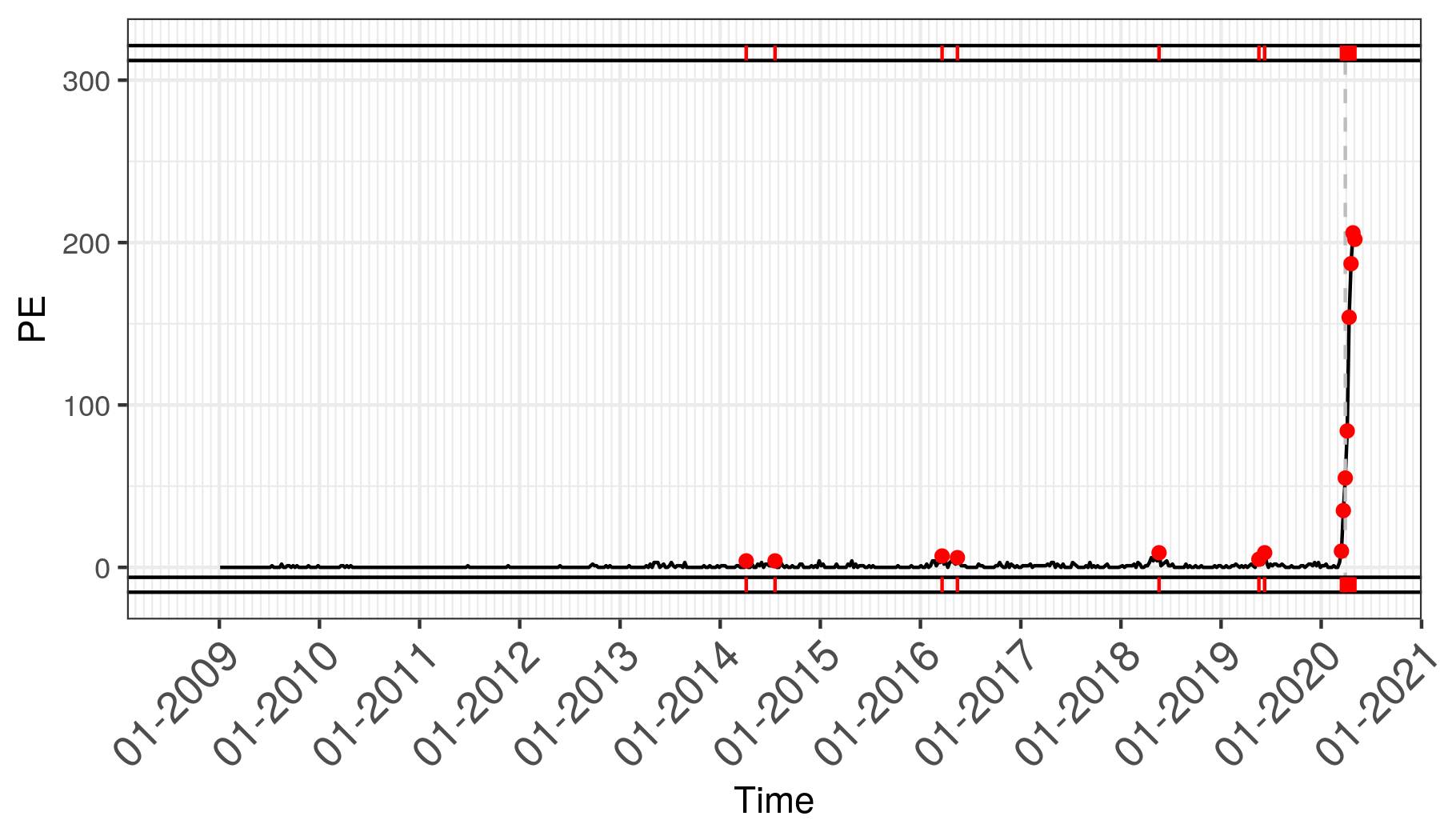}
	}
	\subfloat[Bahia deaths]{
		\label{deaths_BA}
		\includegraphics[width=0.42\textwidth]{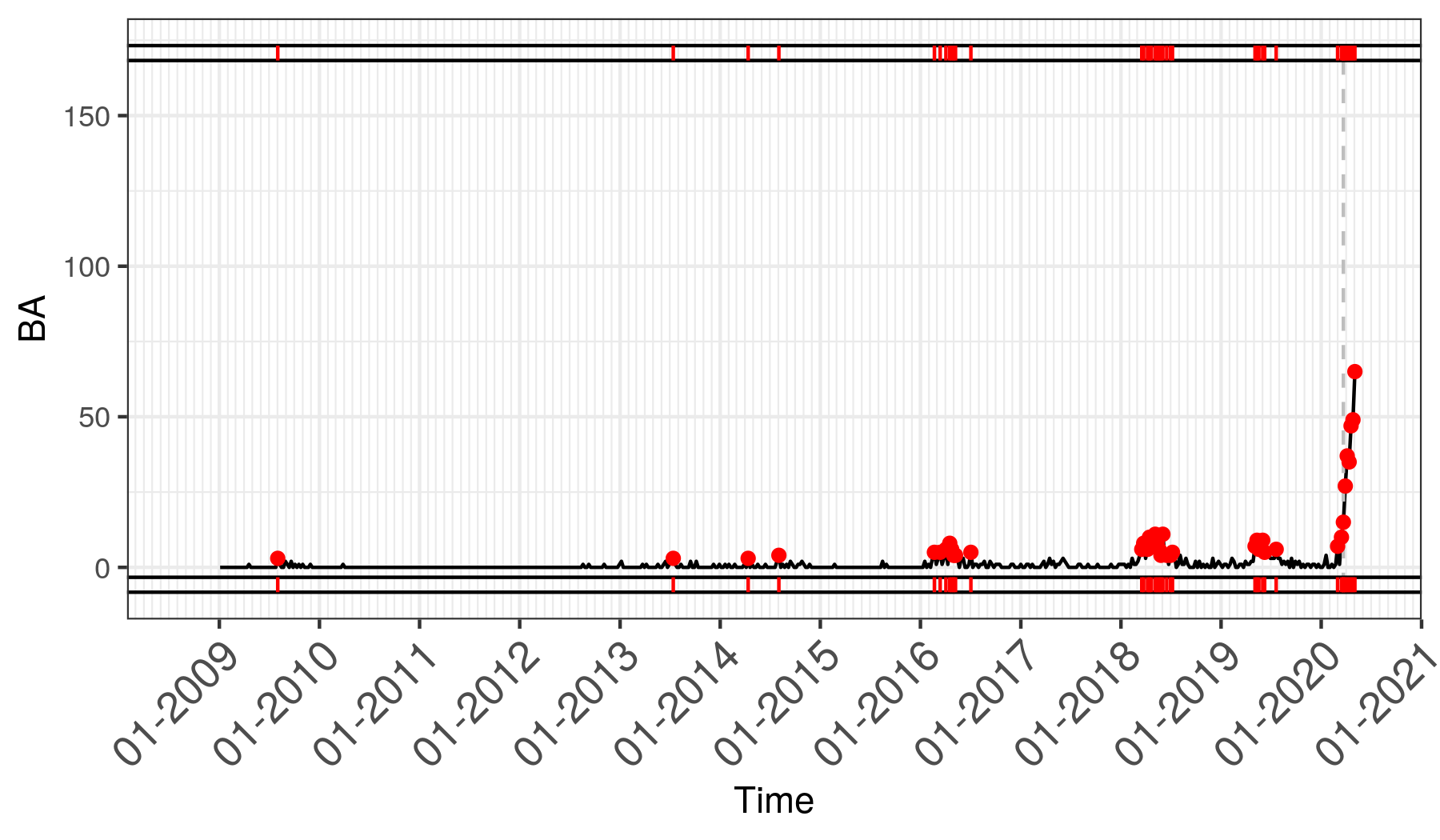}
	}
	
	\subfloat[Distrito Federal deaths]{
		\label{deaths_DF}
		\includegraphics[width=0.42\textwidth]{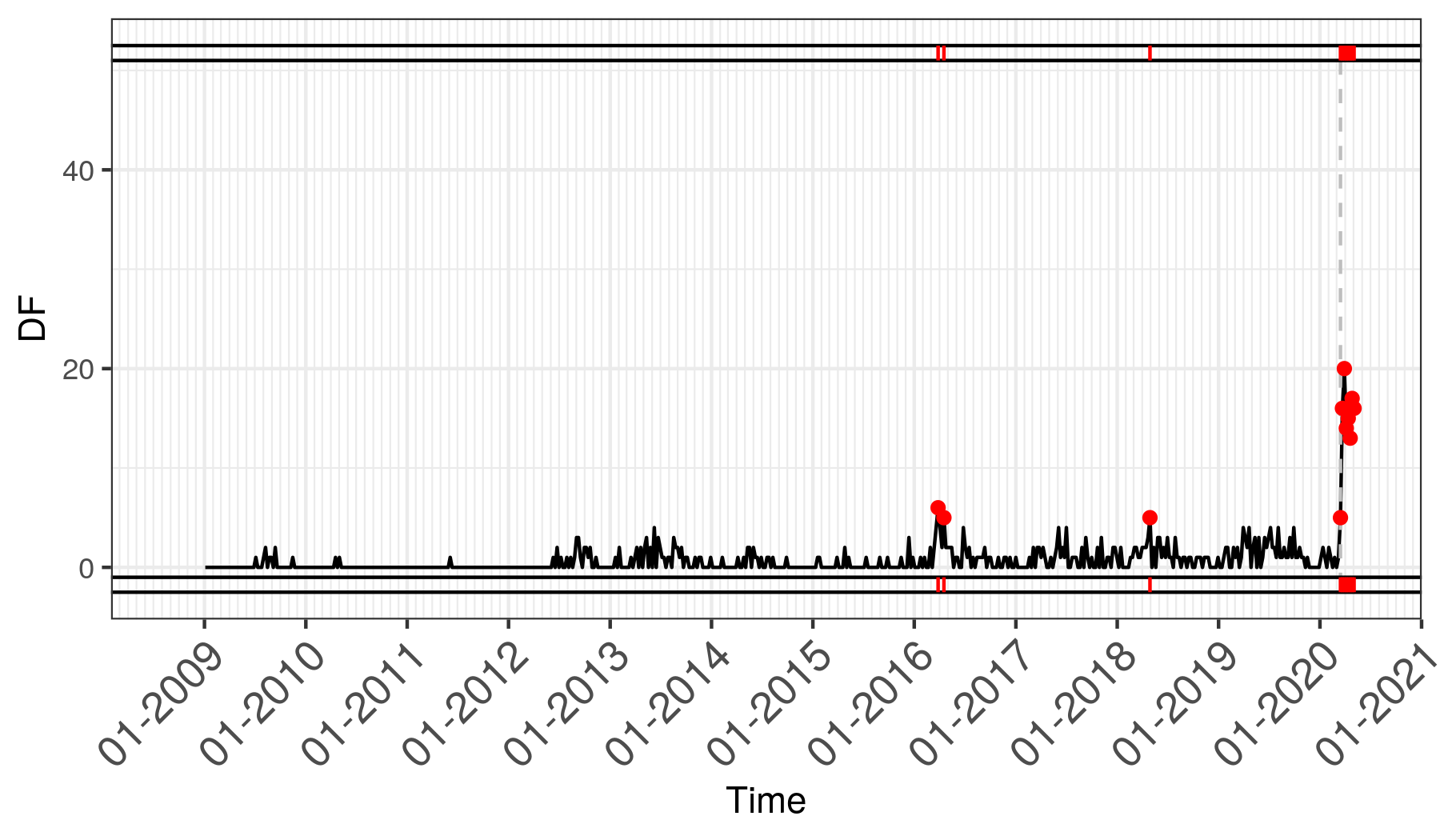}
	}
	\subfloat[São Paulo deaths]{
		\label{deaths_SP}
		\includegraphics[width=0.42\textwidth]{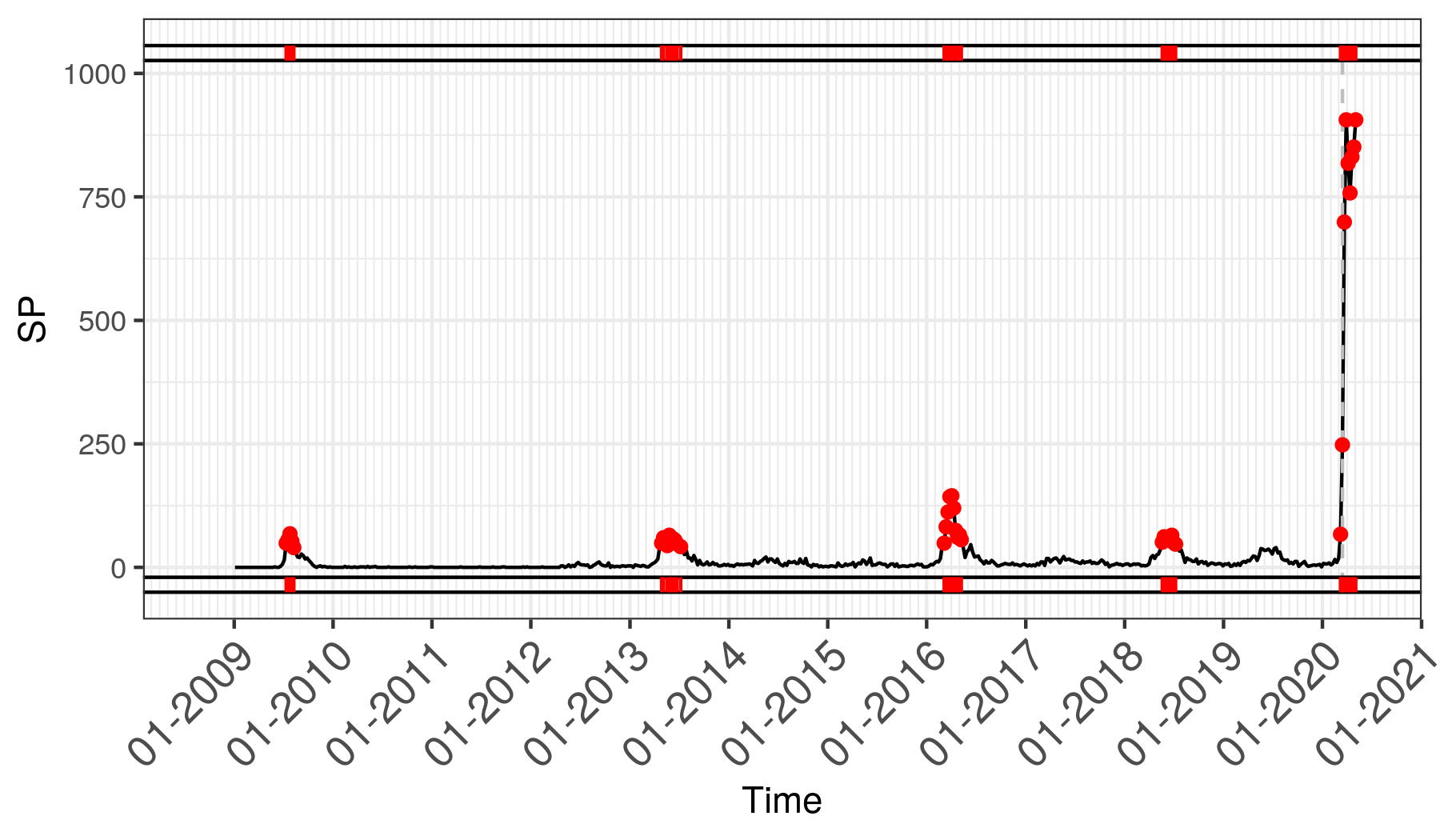}
	}
	
	\subfloat[Rio de Janeiro deaths]{
		\label{deaths_RJ}
		\includegraphics[width=0.42\textwidth]{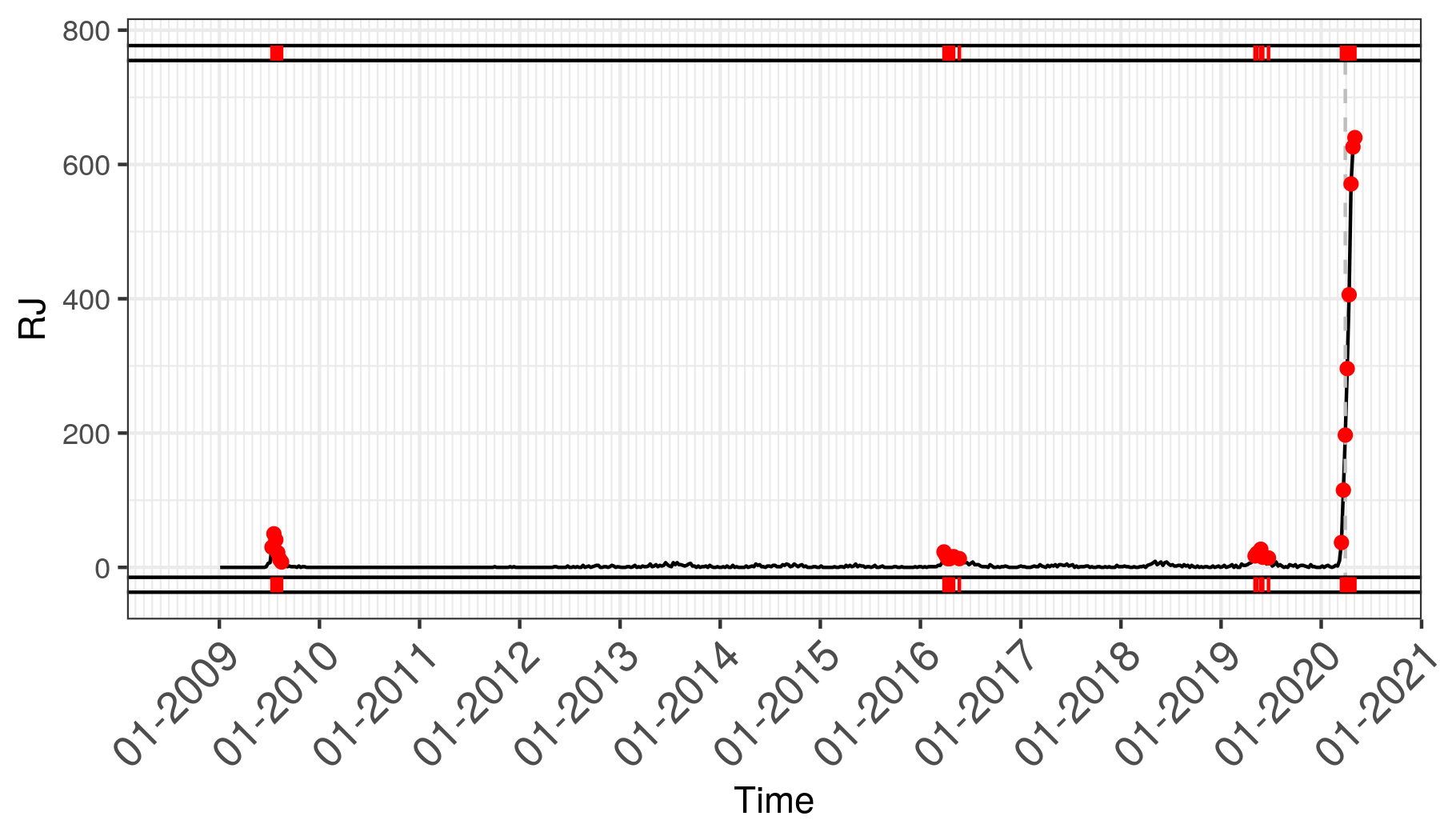}
	}
	\subfloat[Minas Gerais deaths]{
		\label{deaths_MG}
		\includegraphics[width=0.42\textwidth]{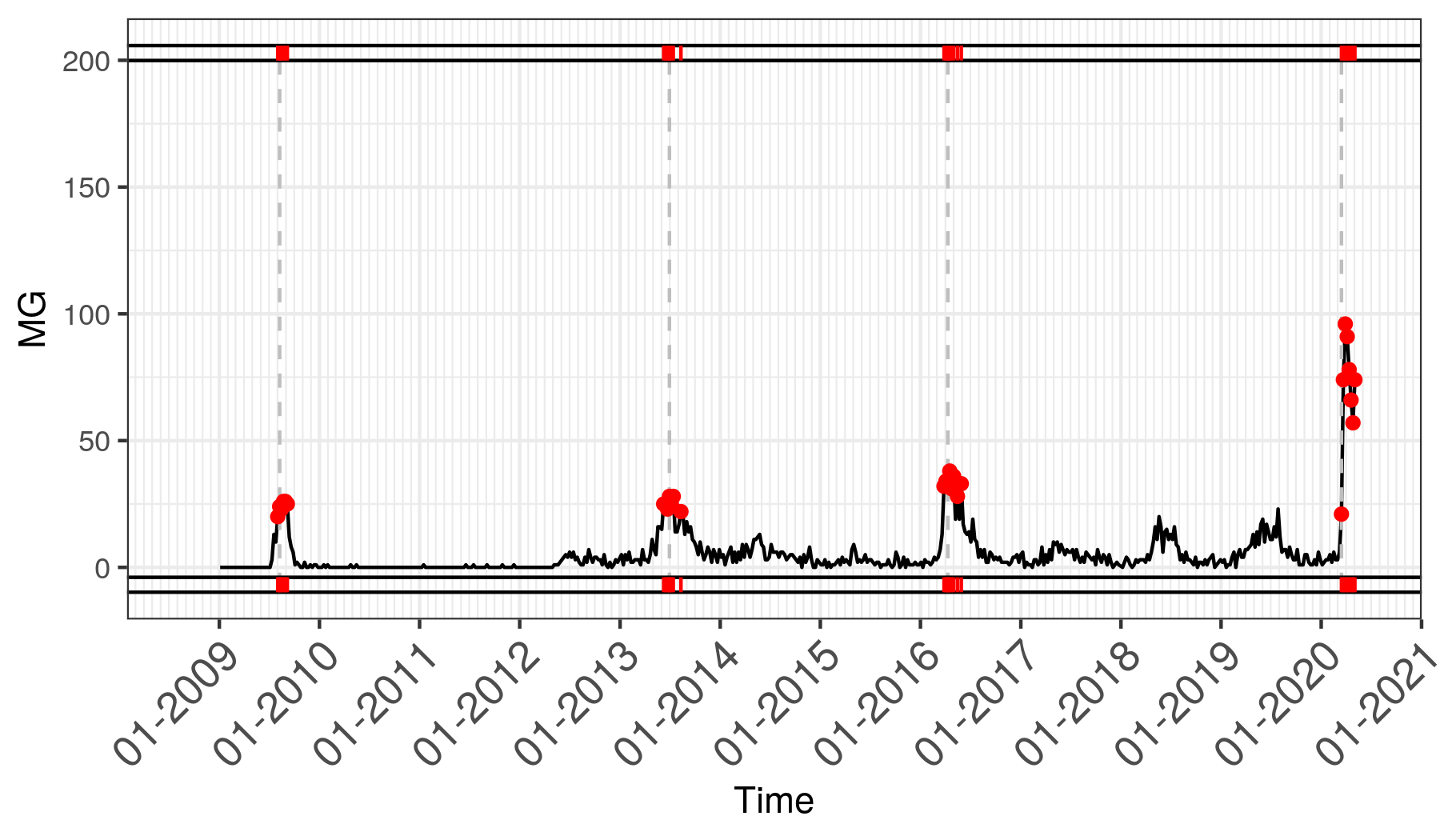}
	}
	
	\subfloat[Paraná deaths]{
		\label{deaths_PR}
		\includegraphics[width=0.42\textwidth]{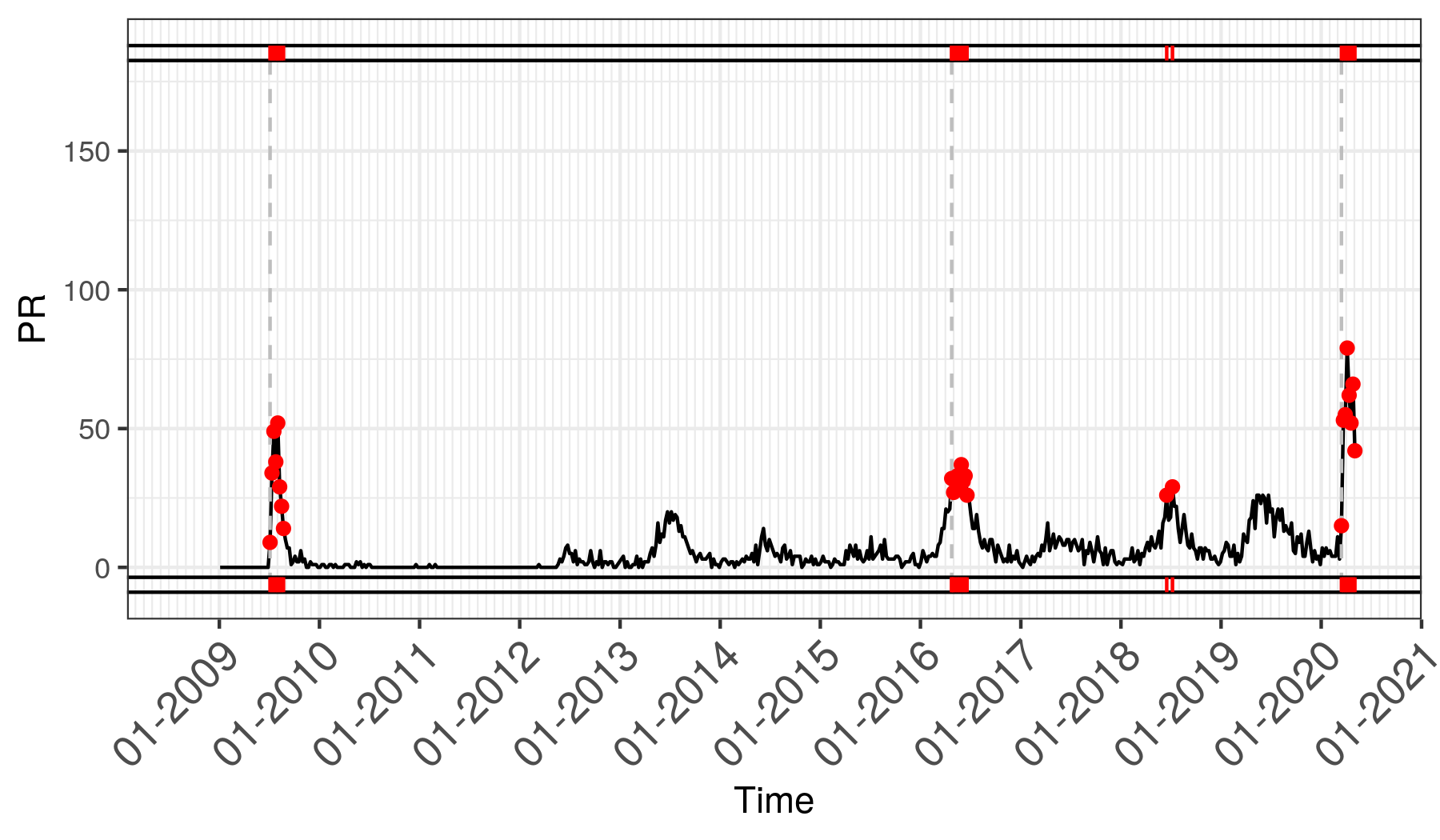}
	}
	\subfloat[Rio Grande do Sul deaths]{
		\label{deaths_RS}
		\includegraphics[width=0.42\textwidth]{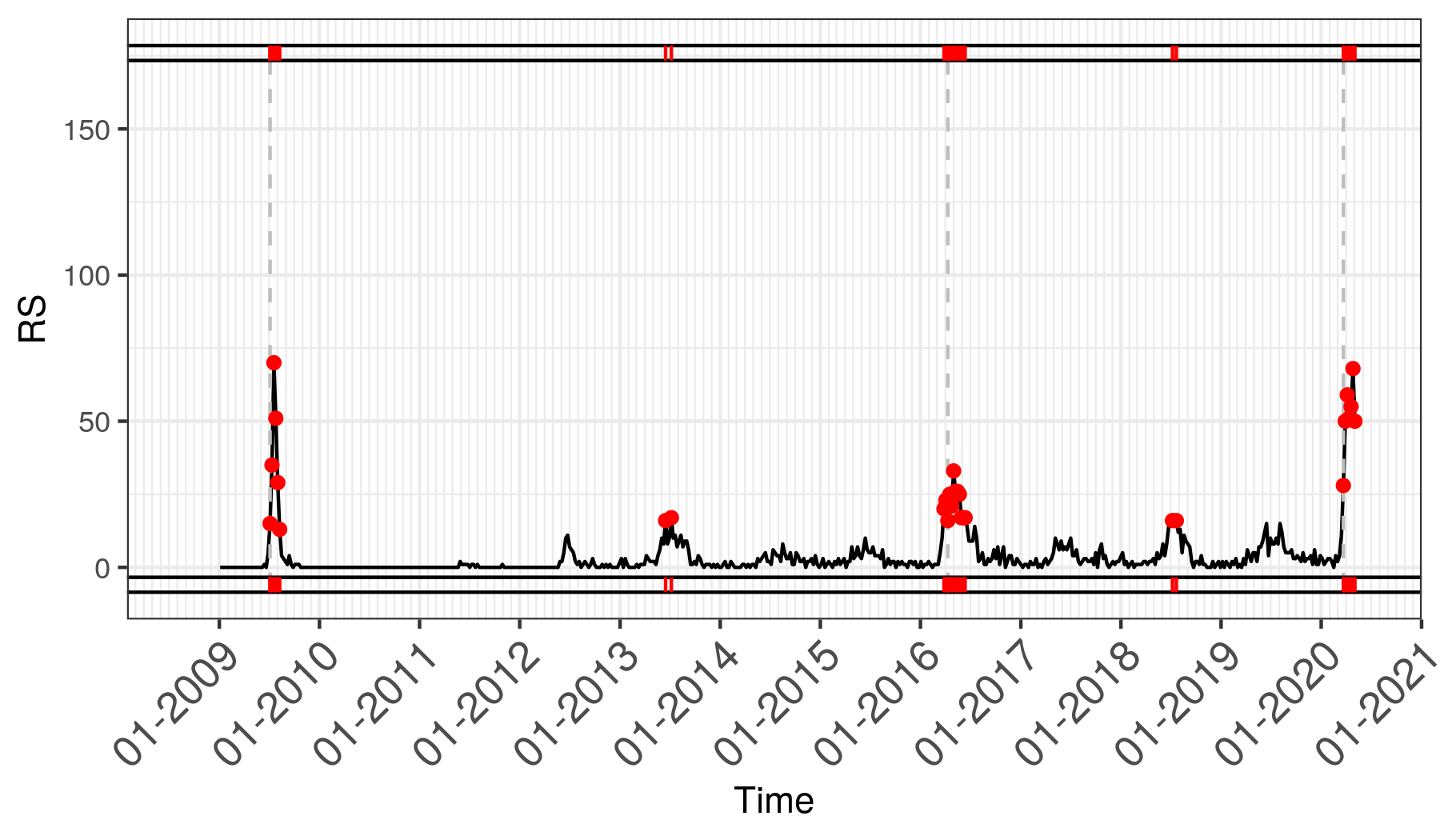}
	}
	\caption{Event detection in time series of deaths}
	\label{fig_deaths}
\end{figure}

The detection of change points and anomalies in the time series of SARI hospitalization in Brazil was an important aspect to understand the beginning process of the pandemic situation of COVID-19 in the country. It also enabled the analyses of epidemic moments over the last years. In Figures \ref{fig_cases} and \ref{fig_deaths}, it is possible to observe the behavior of data and specificity of the most affected Brazilian state\footnote{The graphics for all states are available at \url{https://eic.cefet-rj.br/~dal/covid-19-under-report/}}.

Amazonas state is the epidemic center in the North region, and its capital, Manaus, was the first capital from Brazil to suffer from a wave of deaths. The state presented in 2019 an increase in the number of hospitalizations. This increase is also observed in other states from 2016 until 2019. The Amazonas time series shows some anomalies, but just one change point for both the number of cases (Figure \ref{cases_AM}) and deaths (Figure \ref{deaths_AM}). The change point in the number of deaths and cases is marked, respectively, in the last week of March 2020 and one week later, which corresponds to the thirteenth and fourteenth epidemiological weeks.

In the Northeast region, it is possible to highlight the cases and deaths that occurred at Ceará (Figures \ref{cases_CE} and \ref{deaths_CE}), Pernambuco (Figures \ref{cases_PE} and \ref{deaths_PE}), and Bahia (Figures \ref{cases_BA} and \ref{deaths_BA}). Both Ceará and Pernambuco displayed the highest numbers in the region. The Ceará state shows the same behavior as Amazonas, presenting the change points in the thirteenth and fourteenth weeks. Meanwhile, in Pernambuco, both deaths and cases occurred one week early. In Bahia and Pernambuco, the number of cases and deaths show, between 2016 and 2019, a similar increase and decrease in shaping a curve between March and July.

Distrito Federal, located in the central-West region of Brazil, was then considered one of the main focuses of COVID-19 contagion beside Rio de Janeiro and São Paulo. The peak of the number of cases (Figure \ref{cases_DF}) in Distrito Federal is in August of 2009, during the H1N1 epidemic. However, the number of deaths (Figure \ref{deaths_DF}) caused by H1N1 was not as expressive as the numbers registered by COVID-19.

The Southeast is the most populous region and the most infected area in the country. São Paulo was the first state to register a case and death by COVID-19. They, respectively, occurred in February and March. It is still the epicenter of the disease in Brazil. The state has the mark of the change point for cases (Figure \ref{cases_SP}) and deaths (Figure \ref{deaths_SP}) at the eleventh epidemiological week. It quickly reached the highest registered numbers, more than 4000 cases and 800 deaths in a week.

Rio de Janeiro, also a southeast region, was impacted by SARS-CoV-2. It is possible to observe in the cases (Figure \ref{cases_RJ}) two change points. The first one is 2009 and the second in 2020. However, the number of observed change points for the number of deaths (Figure \ref{deaths_RJ}) occurred only once, in 2020, showing the seriousness of this pandemic. 

Another southern state is Minas Gerais. It registered outliers in 2015 and more stable behavior between 2017 and 2019 for the numbers of cases (Figure \ref{cases_MG}) and deaths (Figure \ref{deaths_MG}). In 2020 the method used detected the change point in the same epidemiological week not only for cases but also for the number of deaths.

The southern states were also impacted by the 2009 H1N1 crisis. According to the time series it is noticeable that Paraná and the Rio Grande do Sul were affected in the number of cases (respectively Figures \ref{cases_PR} and \ref{cases_RS}). On the other hand, if we compare the number of deaths, we can observe and analyze the lethality between these two epidemic moments. Paraná is an example of that analysis, where the maximum point of cases in 2009 surpasses 5,000. Meanwhile, the top of 2020 cases (until the current moment) is less than 1,000. Nonetheless, when observing the number of deaths (Figure \ref{deaths_PR}), the highest numbers occurs in 2020.

\subsection{Under-Reporting Rates} \label{subsec:subn-rates}

The under-reporting rates were computed according to the proposed methodology. Tables \ref{table:rates-cases} and \ref{table:rates-deaths} show the values of the under-reporting rates of cases and deaths for the 27 states of Brazil. In the second column ($cum. novelty$) are the novelty values ($\eta_{i}$) calculated in the methodology. In the third column ($cum. cases$ $DT\_SARI\_c$ and $cum. deaths$ $DT\_SARI\_d$) are the number of cases/deaths classified as SARS-CoV-2 in Infogripe data. In the fifth column ($cum. cases$ $DT\_HM\_c$ and $cum. deaths$ $DT\_HM\_d$) are the number of cases/deaths reported by the Ministry of Health, for comparison purposes. The information published by the Ministry of Health are all confirmed cases/deaths of COVID-19, regardless of whether there was hospitalization for SARI or not, so they capture a broader number of reported records. 

The under-reporting rates presented in this paper can be applied to compute the under-reported cases or deaths of COVID-19 in each state. It is computed by multiplying the under-reporting rates with the number of confirmed cases or deaths of COVID-19. The result can be added to reported cases/deaths to estimate the expected number of cases or deaths of COVID-19 in the state.

\begin{table}[ht!]
	\centering
	\caption{Under-reporting rates of cases of COVID-19 for the states of Brazil}
	\begin{tabular}{C{0.75cm} C{2cm} C{2.0cm} C{2cm} C{2cm}}
		\hline\noalign{\smallskip}
		UF & cum. novelty ($DT\_SARI\_c$) & cum. cases ($DT\_SARI\_c$) & cases rate & cum. cases ($DT\_HM\_c$) \\
		\hline\noalign{\smallskip}
		AC	&	0	&	 13&	- $^{\circ}$ 	&	553	\\
		AL	&	308	&	 152&	1.026 $\pm$ 0.026	&	1372	\\
		AM	&	3824& 2165&	0.766 $\pm$ 0.018	&	6062	\\
		AP	&	83	&	 39&	1.128 $\pm$ 0.026	&	1187	\\
		BA	&	832	&	 350&	1.377 $\pm$ 0.071	&	3267	\\
		CE	&	4704& 2085&	1.256 $\pm$ 0.015   	&	8231	\\
		DF	&	401	&	 251&	0.598 $\pm$ 0.064	&	1566	\\
		ES	&	243	&	 152&	0.599 $\pm$ 0.086	&	2948	\\
		GO	&	363	&	 162&	1.241 $\pm$ 0.191	&	825	\\
		MA	&	650	&	 132&	3.924 $\pm$ 0.030	&	3805	\\
		MG	&	3553&	 484&	6.341 $\pm$ 0.024	&	2023	\\
		MS	&	420	&	 53&	6.925 $\pm$ 0.110	&	266	\\
		MT	&	360	&	 85&	3.235 $\pm$ 0.071	&	331	\\
		PA	&	1390&	 909&	0.529 $\pm$ 0.017	&	3460	\\
		PB	&	619	&	 168&	2.685 $\pm$ 0.030	&	1034	\\
		PE	&	3158&	 976&	2.236 $\pm$ 0.018	&	8145	\\
		PI	&	602	&	 186&	2.237 $\pm$ 0.048	&	665	\\
		PR	&	1779&	 389&	3.573 $\pm$ 0.136	&	1492	\\
		RJ	&	8069&	3679&	1.193 $\pm$ 0.009	&	10546	\\
		RN	&	386	&	 207&	0.865 $\pm$ 0.024	&	1366	\\
		RO	&	27	&	 15&	- $^{\circ}$    	&	653	\\
		RR	&	71	&	 45&	0.578 $\pm$ 0.022	&	668	\\
		RS	&	2175&	 615&	2.537 $\pm$ 0.093	&	1619	\\
		SC	&	972	&	 303&	2.208 $\pm$ 0.096	&	2346	\\
		SE	&	92	&	 62&	0.484 $\pm$ 0.065	&	601	\\
		SP	&	25938 & 13057&	0.987 $\pm$ 0.025	&	31174	\\
		TO	&	141	&	 38&	2.711 $\pm$ 0.053	&	191	\\
		\hline\noalign{\smallskip}
	\end{tabular}
	
	$^{\circ}$ The difference between computed novelty and random noise was not statistically significant.
	
	\label{table:rates-cases}
\end{table}

\begin{table}[ht!]
	\centering
	\caption{Under-reporting rates of deaths by COVID-19 for the states of Brazil}
	\begin{tabular}{C{0.75cm} C{2cm} C{2.0cm} C{2cm} C{2cm}}
		\hline\noalign{\smallskip}
		UF & cum. novelty ($DT\_SARI\_d$) & cum. deaths ($DT\_SARI\_d$) & death rate & cum. deaths ($DT\_HM\_d$)\\
		\hline\noalign{\smallskip}
		AC 	&	0	&	 13&	 - $^{\circ}$ 	&	21	\\
		AL 	&	49	&	 34&	 0.441 $\pm$ 0.029 	&	58	\\
		AM 	&	2023&	1147&	 0.764 $\pm$ 0.003 	&	501	\\
		AP 	&	26	&	 21&	 - $^{\bullet}$ 	&	40	\\
		BA 	&	200	&	 110&	 0.818 $\pm$ 0.027 	&	123	\\
		CE 	&	1429&	 983&	 0.454 $\pm$ 0.004 	&	614	\\
		DF 	&	90	&	 33&	 1.727 $\pm$ 0.061 	&	31	\\
		ES 	&	91	&	 84&	 - $^{\bullet}$ 	&	102	\\
		GO 	&	61	&	 36&	 - $^{\bullet}$ 	&	30	\\
		MA 	&	60	&	 31&	 0.935 $\pm$ 0.032 	&	224	\\
		MG 	&	434	&	 94&	 3.617 $\pm$ 0.096 	&	88	\\
		MS 	&	26	&	 8&	 - $^{\circ}$ 	&	9	\\
		MT 	&	34	&	 19&	 - $^{\bullet}$ 	&	12	\\
		PA 	&	473	&	 414&	 0.143 $\pm$ 0.005 	&	273	\\
		PB 	&	133	&	 86&	 0.547 $\pm$ 0.023 	&	74	\\
		PE 	&	653	&	 369&	 0.770 $\pm$ 0.005 	&	628	\\
		PI 	&	86	&	 34&	 1.529 $\pm$ 0.059 	&	26	\\
		PR 	&	287	&	 83& 2.458 $\pm$ 0.096 	&	90	\\
		RJ 	&	2236&	1577&	 0.418 $\pm$ 0.003 	&	951	\\
		RN 	&	83	&	 68&	 0.221 $\pm$ 0.029 	&	59	\\
		RO 	&	3	&	 4&	 - $^{\circ}$ 	&	23	\\
		RR 	&	17	&	 16&	 - $^{\bullet}$ 	&	9	\\
		RS 	&	303	&	 77&	 2.935 $\pm$ 0.104 	&	62	\\
		SC 	&	104	&	 48&	 1.167 $\pm$ 0.062 	&	52	\\
		SE 	&	22	&	 13&	 0.692 $\pm$ 0.077 	&	14	\\
		SP 	&	5131&	3207&	 0.600 $\pm$ 0.010 	&	2586\\
		TO 	&	13	&	 16&	 - $^{\circ}$ 	&	4	\\
		\hline\noalign{\smallskip}
	\end{tabular}
	
	$^{\circ}$ The difference between computed novelty and random noise was not statistically significant.
	
	$^{\bullet}$ The difference between computed novelty and reported values was not statistically significant.
	\label{table:rates-deaths}
\end{table}

The under-reporting rates of cases vary between 0.484 and 6.925, while the under-reporting rates of deaths vary between 0.143 and 3.617.
Among the states for which it was possible to calculate the two rates, the majority had higher under-reporting rate of cases than under-reporting rate of deaths. Only the states RS, DF and SE behave differently. DF is highlighted because it has a death rate almost 3 times higher than that of cases.

There is no dominant pattern between states in each region of Brazil. It suggests that under-reporting is a characteristic of each state. The regional similarity is not a relevant factor. The states of MG and MS have the highest rates of under-reporting of cases. The rate of under-reporting of deaths is high in the MG and the RS.

The DF, SP and RJ are identified as the focus of the contagion of COVID-19 in Brazil. Nevertheless, both DF and SP are not the ones with the highest rates of under-reporting. It may be because they might be better structured and less susceptible to reporting failures. This same observation is not valid for the states MS and MG in the same regions (mid-west and southeast regions, respectively), which have the highest rates of under-reporting of cases across Brazil.

The proposed model did not capture under-reporting of cases in the AC and RO or deaths in the states of AC, RO, MS, MT, TO, GO, RR, AP, and ES. These are the cases in which either a novelty cannot be detected ($^{\circ}$) or under-reporting cannot be observed ($^{\bullet}$). MS stands out since, despite having a high-rate of under-reporting of cases (second highest among states), the under-reporting of deaths was not observed.

Regarding the margin of error considered for the case rates, the states of the mid-west and south regions are highlighted. A factor that may have been determinant for this result is their historical temperature. As they have low temperatures, they generally, a higher number of SARI records. Thus, the novelty modeled in this work takes longer to be noticed, as it needs to reach even higher values to provide statistically significant changes.

\section{Final Remarks} \label{sec:final_remarks}

This study aimed to estimate the rates of under-reporting of cases and deaths in the states of Brazil. The methodology is based on the concepts of inertia and the use of event detection techniques to study the time series of hospitalized SARI cases. All methods and parameters used in the methodology are justified, based on the modeling or available data.

We introduced the concept of novelty about SARI analysis to observe the under-reporting of COVID-19. Consequently, COVID-19 causes a rupture in the SARI series inertial behavior, changing the statistical properties of the time series. This break is identified by event detection techniques. If the change occurred is due to COVID-19, the computed novelty then corresponds to estimates of the values of cases and deaths from the disease. From this, under-reporting rates were computed.

Since the under-reporting is inferred from SARI data, estimates are limited to cases of COVID-19 that manifested specific symptoms (fever, cough or sore throat, dyspnea or oxygen saturation below $95\%$ and difficulty to breathe) and were hospitalized. It corresponds to a portion of the cases of COVID-19, as many individuals have milder symptoms or are even asymptomatic. Thus, we can consider the computed of under-reporting rates as very conservative since it only considers symptomatic and hospitalized cases of the disease.

For this same reason, we believe that the results are better characterized for deaths than for cases, since people who died are much more likely to have been hospitalized and, therefore, present in SARI data. This is quite clear when looking at the Tables \ref{table:rates-cases} and \ref{table:rates-deaths}. While in the table of cases (Table \ref{table:rates-cases}) the data from the Ministry of Health mostly account for many more cases than those determined in the novelty, in the Table of deaths (Table \ref{table:rates-deaths}) the number of deaths found of the novelty are higher.

Limitations should be noted. One limitation is inherent to the dataset used. In times of epidemic, health services tend to be more sensitive and report more occurrences. Thus, the increase in the number of SARI cases in 2020 is partially justified by the over-notification of health units. This super notification, however, is mitigated when only hospitalized cases are observed.
Another limitation is due to random noise $\epsilon_i$. The states in which $\epsilon_i$ were higher are slower to characterize the novelty $\eta_i$. Again, the computed under-reporting rates presented in this paper are conservative. They can be improved by predicting $\epsilon_i$ using autoregressive models.

\section*{Acknowledgments}

The authors thank CNPq, CAPES (finance code 001), and FAPERJ for partially funding this research.

\bibliographystyle{abbrv}
\bibliography{references}

\end{document}